\documentclass[sn-basic]{sn-jnl}

\usepackage{graphics}
\usepackage{graphicx}%
\usepackage{multirow}%
\usepackage{amsmath,amssymb,amsfonts}%
\usepackage{amsthm}%
\usepackage{mathrsfs}%
\usepackage[title]{appendix}%
\usepackage{xcolor}%
\usepackage{textcomp}%
\usepackage{manyfoot}%
\usepackage{booktabs}%
\usepackage{algorithm}%
\usepackage{algorithmicx}%
\usepackage{algpseudocode}%
\usepackage{listings}%

\usepackage{hyperref}

\usepackage[utf8]{inputenc}
\usepackage[braket, qm]{qcircuit}

\usepackage[caption=false]{subfig}

\usepackage{listings}

\usepackage{amsmath,amssymb,amsfonts}
\usepackage{graphicx}
\usepackage{textcomp}

\theoremstyle{thmstyleone}%
\theoremstyle{thmstyletwo}%

\theoremstyle{thmstylethree}%

\usepackage{makecell}

\begin{document}

\title[Article Title]{Quantum Convolutional Neural Networks with Interaction Layers for Classification of Classical Data}


\author[1]{\fnm{Jishnu} \sur{Mahmud}}\email{jishnu.mahmud@ieee.org}

\author[1]{\fnm{Raisa} \sur{Mashtura}}\email{raisa.mashtura@ieee.org}

\author*[1]{\fnm{Shaikh Anowarul} \sur{Fattah}}\email{fattah@eee.buet.ac.bd}

\author[2]{\fnm{Mohammad} \sur{Saquib}}\email{saquib@utdallas.edu}



\affil[1]{\orgdiv{Department of Electrical \& Electronic Engineering}, \orgname{Bangladesh University of Engineering \& Technology}, \orgaddress{\city{Dhaka}, \postcode{1000}, \country{Bangladesh}}}

\affil[2]{\orgdiv{Department of Electrical Engineering}, \orgname{The University of Texas at Dallas}, \orgaddress{\city{Texas}, \postcode{75083-0688}, \country{USA}}}


\abstract{Quantum Machine Learning (QML) has come into the limelight due to the exceptional computational abilities of quantum computers. With the promises of near error-free quantum computers in the not-so-distant future, it is important that the effect of multi-qubit interactions on quantum neural networks is studied extensively. This paper introduces a Quantum Convolutional Network with novel Interaction layers exploiting three-qubit interactions, while studying the network's expressibility and entangling capability, for classifying both image and one-dimensional data. The proposed approach is tested on three publicly available datasets namely \textit{MNIST}, \textit{Fashion MNIST}, and \textit{Iris} datasets, flexible in performing binary and multiclass classifications, and is found to supersede the performance of existing state-of-the-art methods. }


\keywords{Quantum Machine Learning, classification, entanglement, quantum gates, qubits.}



\maketitle

\section{Introduction}\label{sec1}
{I}{n} this era of artificial intelligence, a constant improvement in computation speed, accuracy, and precision is a necessity. This widespread success in the world of computing over the last decade can be attributed to both the development of efficient software algorithms and the advancements in computational hardware. However, the physical limits of semiconductor fabrication in the post-Moore's Law era raise concerns about the extrapolation of its effectiveness in the future. On the other hand, significant advancements have been made in the field of quantum computing, which has shown promise as a potential solution for modern computing problems. Quantum computing exploits the laws of quantum mechanics to store and process information in quantum devices (Online Resource 1), using qubits instead of classical bits, which enables them to solve problems intractable for classical computers (\cite{r1}). 
The era of quantum computing, currently referred to as the Noisy Intermediate Scale Quantum (NISQ) era, is characterized by the lack of absolute control over the qubits due to errors arising from quantum decoherence, crosstalk, and imperfect calibration, thereby limiting the number of qubits used on quantum computers. However, the revelation in January 2022 that quantum computing in silicon hit 99\% fidelity (\cite{mkadzik2022precision}) indicates a more significant similarity between the desired and actual quantum states. This result promises near-error-free quantum computing and indicates that they are close to being utilized in large-scale applications, further motivating the development of various machine learning algorithms to be implemented on quantum devices.

A quantum circuit proposed in this work is designed in the spirit of QCNN structure and possesses minimal trainable parameters. The robustness of QCNNs against the “barren plateau” issue is expected to be exhibited by the proposed network. The rapid advancement of quantum hardware indicates that considerably more intricate operations on qubits will soon be possible. Several works in the broader spectrum of Quantum Computing share this objective for quantum machine intelligence research (\cite{nguyen2023biomarker}, \cite{ayoade2022artificial}).  Although achieving three-qubit interactions on current technology is practically challenging, this paper investigates the comparative advantage in a network's performance due to the addition of such layers. It must be noted that the number of trainable parameters and the total number of qubits have been kept to a minimum such that they can be implemented on NISQ devices for the purpose of comparison with other methods. The paradox of deploying three-qubit interactions on the network that is, at its core, intended to operate on NISQ devices is acknowledged by the authors. When moving from NISQ to more potent quantum computers, more qubits would be used, the circuit depth would expand, and there would be more qubit interactions with multi-qubit gates. This study evaluates how expanded multi-qubit interaction improves QCNN network performance compared to networks limited to two-qubit operations by keeping a small number of qubits and trainable parameters. 

The rest of the paper is arranged in the following way: Section $2$ highlights the current state of QML literature, discussing related works along with a brief overview of the objective of this work in the current scheme of this field. Section $3$ discusses the details of the proposed architecture, which is divided into three subdivisions corresponding to the three main subsystems. Section $4$ focuses on simulation and results, describing the different datasets, configurations, and parameters used to benchmark the network. The resulting accuracies and costs are also reported. Finally, in section $5$, conclusions are drawn, and the scope of related future works is discussed.

\section{Related Works}
In the early works of Quantum Machine Learning (QML), the power of quantum algorithms is used to solve various subtasks as clever modifications to the already existing classical machine learning algorithms with the goal of increased efficiency and speedup (\cite{schuld2022quantum}). These tasks involve various mathematical and algorithmic processes that are direct results developed from the fundamentals of traditional quantum computing (\cite{Q_algo_fitting}), (\cite{kerenidis2022quantum}).
The NISQ era has given rise to another genre of work where QML has been implemented, in its true sense, on variational quantum circuits for various applications involving classical data. The process involves designing a quantum circuit with free parameters, which are iteratively updated using gradient descent by minimizing an objective cost function. The cost functions in most of these works are classical and, therefore, similar to the ones used in current machine learning literature. This area of QML deals with designing a variational quantum network, choosing objective cost functions, and experimenting with the network’s trainability, expressibility, and ability to be generalized into various applications (\cite{farhi2018classification}) (\cite{ckt_centric}). The NISQ era is characterized by the exponential difficulty of implementing and simulating such quantum networks as the number of qubits, circuit depth, and inter-qubit interactions increase, causing multiple researchers to resort to designing smaller networks to solve a scaled-down version of a real-world task.
QML has already been implemented to address one of the most fundamental machine learning problems, i.e., classification. \cite{mengoni_kernel} has reviewed and summarized the mathematical basis for various kernel-based QML algorithms widely used for the task. One such algorithm that has been reviewed is the Quantum Support Vector Machine (QSVM) initially proposed in \cite{QSVM_rebentrost}. The work on QSVM shows that the kernel-based quantum binary classifier has complexity logarithmic to the size of the feature space and the number of training samples. This means that the classical Support Vector Machine (SVM) model can be solved in a run-time proportional to $O(log(\epsilon^{-1})poly(N, M))$ (\cite{boyd2004convex}). In contrast, QSVM is shown to have a run-time of $O(Log(NM))$ where $N$ is the dimension of feature space, $M$ is the number of training samples, and $\epsilon$ is the accuracy. Although such works lay the mathematical foundation of various QML models, with claims of quantum advantages, benchmarking them on real-life datasets and implementing them on real quantum hardware in the NISQ era is a separate challenge. This task of experimentation and tweaking of such algorithms to enhance their performance in various fields has been taken up in a handful of papers for a wide range of applications such as in medicine (\cite{lung_cancer}), weather forecasting (\cite{weather}), quantum chemistry (\cite{chemical}), and many more. The networks proposed in these papers tackle reasonably low complexity problems displaying marginal quantum advantages over their classical counterpart.

The search for devising various convolutional networks in the quantum domain is introduced in \cite{cong_quantum_org}, where the concept of quantum convolutional neural networks (QCNNs) is proposed. Their architecture also claims to tackle the exponential complexity of many-body problems, making them suitable for use in quantum physics problems. Advancing the field of QCNNs, a parameterized quantum neural network (QNN) with an enhanced feature mapping process has been designed in \cite{liu_the2nd}. Their proposed network is called a quantum-classical CNN (QCCNN), suitable for NISQ devices. Several quantum counterparts of various classical machine learning models have been proposed over recent years, claiming superior performances in various categories, such as accuracy and speed. However, these are observed to tackle a subset of a particular real-world problem. 

Ever since the proposal of QCNNs and the availability of quantum simulators and quantum computers, much attention has been drawn to devising various methods to improve the performance of classification problems using quantum networks that are implementable on NISQ devices. This is driven by QCNN models being immune to barren plateau problems (\cite{pesah2021absence}) contrary to other structures. The architecture proposed in \cite{tak_hur_boss} has been benchmarked for binary image classification on the \textit{MNIST} and \textit{Fashion MNIST} datasets. A multiclass classification method using a quantum network is also reported in \cite{chalumuri}, which has been proven to perform well on 1D data such as the \textit{Iris dataset}. All these prior studies confirm that a QNN aids speed with a significantly lower number of parameters with better accuracy than their existing classical counterparts using a comparable number of training parameters. However, a crucial aspect of designing parameterized quantum circuits is maintaining sufficient expressibility and entangling capability while keeping it cost-effective \cite{sim2019expressibility}. 


The cost of a quantum circuit is judged by the number of layers and, hence, its parameters, as well as the number of two-qubit operations. This paper explores the relative changes in the performance of a newly proposed QCNN network, which includes limited three-qubit interactions while maintaining a relatively low depth and a small number of parameters for comparability. Although three-qubit gates are practically difficult to implement on NISQ devices and their synthesis using 1 and 2-qubit interactions exponentially increases depth, it is essential to explore the comparative changes in the performance of a network resulting from their addition, which is expected to be a reality considering the rapid growth in the performance of quantum hardware in recent times. These three-qubit interactions are brought forward using novel Interaction Layers in the proposed network, which use a minimal number of trainable parameters. Furthermore, to explore the performance of the proposed network, an ancilla-based classifier is used as the final layer of the circuit to carry out binary and multiclass classification tasks. Considering the current era of quantum computing, it is evident that the display of quantum superiority over its classical counterparts has resorted to small-scaled versions of problems. A concern, as correctly pointed out in \cite{schuld2022quantum}, is that much of the QML literature has been focused on a biased subset of models and conditions that have been aimed toward a display of quantum-enhanced speedup compared to their classical counterparts and have, subsequently, prevented research which delves into a search for systems which are actually effective as quantum models.
On the other hand, although it is difficult to empirically display the quantum advantages of quantum machine learning in the domains of the most complex problems that are currently in the realms of classical machine learning, even the most skeptical extrapolation of QML’s powers from the current research goes on to show a clear expectation of superiority in performance in the near future. Alongside the various enhancements that QML inherits from the domain of quantum computing due to the very nature of quantum systems, QML has some attributes that highlight its superiority in terms of Machine Intelligence. The classical data in QML networks are encoded to an electronic wavefunction via different encoding techniques. These wavefunctions are then subjected to a sequence of quantum gates (unitary operations) to modulate them to their desired states.  These electronic wavefunctions, which represent the states of the qubits, are denoted with complex-valued vectors that can be interpreted as high-dimensional vectors, much like Capsule Networks. Therefore, QML models can be expected to inherit much of the superiority of capsule networks compared to conventional classical convolutional neural networks (CNN). As stated in (\cite{sabour2017dynamic}), a capsule is a collection of neurons representing various properties such as pose, texture, and velocity in the data. Their work explores constructing high-dimensional vectors that represent the existence and orientation of these property vectors and constructing a hierarchal structure with capsules dynamically routing within them from child to parent capsule. These capsules make predictions, and a parent capsule is activated when multiple predictions agree. This shows another striking similarity with QNNs, where the state of a particular qubit is often modulated by considering the state of its neighboring qubits using controlled gates. These similarities fuel the hope that QNN structures, along with their inherent advantage from the quantum computing domain, will also exhibit the advantages of capsule networks with properties such as less sensitivity to input translations and improved generalization. 
The goal of our research, however, is not to display the quantum superiority of the work but to experiment with the proposed network with the novel interaction layers and a three-sub-system structure equipped with the QCNN structure and the ancilla classifier.  The results obtained from this work are compared to other QML models to highlight the relative performance of the network compared to the current QML literature.


\

The major contributions of this work are summarized as follows: 

\

1. A new QCNN architecture is proposed, which is tested with \textit{Amplitude} and \textit{Angle Encoding} schemes separately, considering two different data reduction and encoding techniques. 

\

2. Novel Interaction Layers with three-qubit interactions are introduced, which exhibit sufficient expressibility and exploit the entanglement property of qubits further to help the quantum network learn more nuanced information from the data.

\ 

3. A classifier layer involving ancilla qubits and \textit{CNOT} gates are cascaded with the quantum convolutional structure to accommodate both binary and multiclass classifications.

\ 

4. A unique data aggregation method is used with a combination of measuring the qubits on the expectation values of the Pauli-Z operator and passing the results through a Softmax function.

\

5. The proposed network is tested on three publicly available datasets for binary and multiclass classification, and it is seen that the performance supersedes that of the existing state-of-the-art models using a similar number of parameters. The versatility of the network is further demonstrated in its ability to perform equally well in both image and 1-dimensional data.

\
The use of the proposed ancilla-classifier, along with the use of Interaction Layers in the QCNN structure, is a first to the best of our knowledge.

\section{Proposed Architecture}
\begin{figure*}[t]
    \centering
    \includegraphics[width = \linewidth]{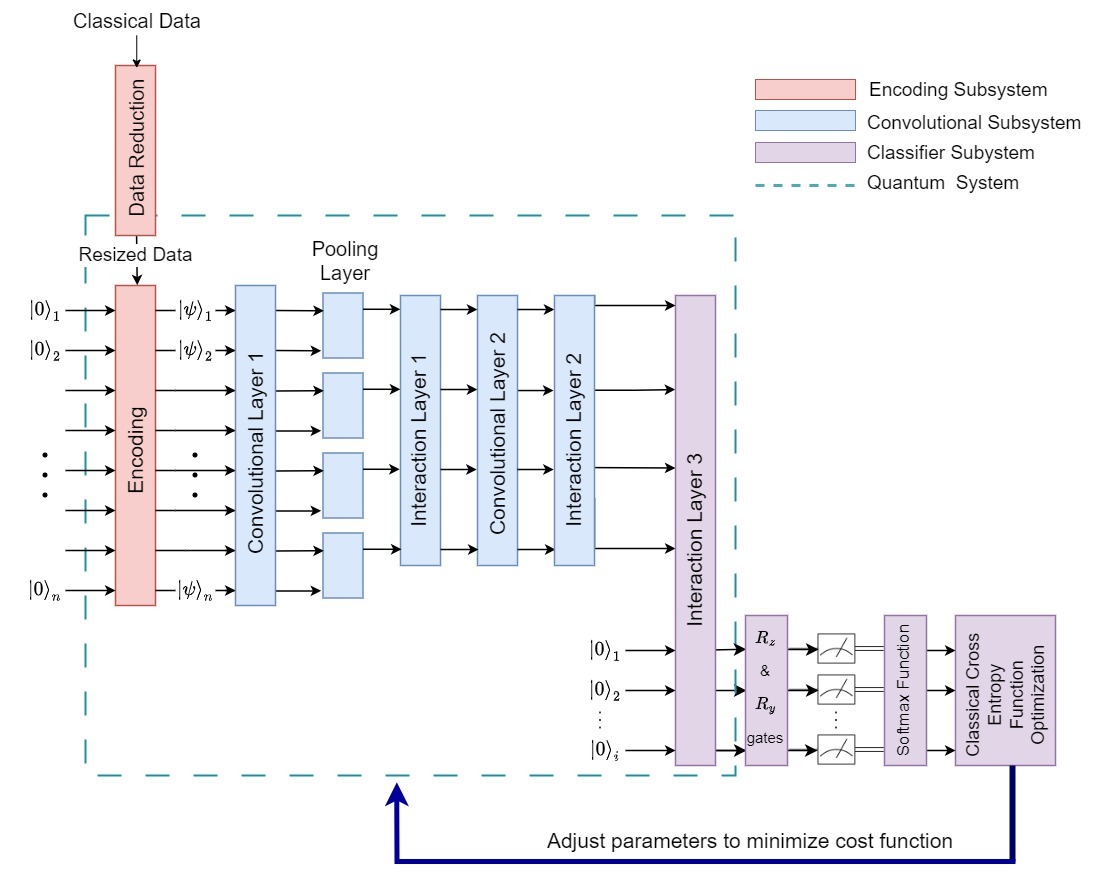}
    \caption{Simplified Block diagram of the proposed architecture showing the Encoding Subsystem followed by the Convolutional Subsystem followed by the Classifier Subsystem. The Quantum System of the architecture comprises quantum gates with trainable parameters, which are optimized classically by minimizing the Cross Entropy loss function. Classical data are embedded on qubits initialized as \ket{0} at the Encoding subsystem of the network. Ancilla qubits with \(i=\) number of classes are used in the Classifier Subystem.  }
    \label{fig:overall_network}
\end{figure*}

The overall network proposed in this work is depicted in a simplified block diagram in Fig. \ref{fig:overall_network}. There are two distinct types of quantum systems used in current literature: discrete-based quantum systems and continuous-based quantum systems. Continuous-based quantum systems are based on the creation of continuous quantum variables, which means that they are vectors in an infinite-dimensional Hilbert Space. On the other hand, discrete-based quantum systems use qubits, which are two-dimensional quantum systems. Although the continuous-based quantum system has a more enriched Hilbert space, the requirement for resources is enormous, and there are considerable difficulties in designing universal gates suited to the system. Therefore, the proposed model in this paper is designed to function as a discrete-based quantum system that still possesses the key attributes of a quantum system while having a more straightforward gate implementation and is better suited to be implemented on NISQ hardware.

The network has been designed in the light of QCNN and is expected to inherit specific advantageous attributes, such as the resistance of such models to the barren plateau problem (\cite{pesah2021absence}). The model architecture can be divided and classified into three subsystems in accordance with their function, each of which has been explained in detail in the subsequent sections. 
In broader terms, the first system is the Data Encoding Subsystem, which is responsible for the preparation of the electronic qubit states representing the classical data. These states are then passed onto the second subsystem, the Convolutional Subsystem, which, via Convolutional, Pooling, and novel Interaction Layers, aims to reduce the number of qubits and modulate the remaining qubit states to highlight the differences between the data classes of the classification problem. The final system, the Classifier Subsystem, uses Interaction layer three as an ansatz to classify the data into one of the classes from the input qubit states representing the highlighted features of the classical data.

Interaction Layers are introduced in various stages of the proposed quantum architecture and are designed to leverage three-qubit interactions through the use of \textit{Toffoli} and parameterized rotational gates. The implementation of these  Layers in various stages of the network can be observed in Fig. \ref{fig:overall_network}. It differs from the earlier quantum convolutional methods, which relied on the reduction of qubits through sequences of convolutional and pooling layers alone. By keeping a small number of qubits and trainable parameters, this study aims to evaluate how expanded multi-qubit interaction improves QCNN network performance in comparison to networks limited to two-qubit operations.


It is expected that the incorporation of these novel interaction layers will enable the network to extensively span the Hilbert space as well as exploit the entanglement property further for improved classification performance. The use of Toffoli gates, enabling three-qubit interactions in QCNN networks, is a first to the best of our knowledge.

\subsection{The Encoding Subsystem}

\subsubsection{Data Preprocessing}
The number of qubits and, therefore, the size of a QNN is bound by the current limitations of NISQ computing technology, in contrast to classical models, which often possess many trainable parameters due to their substantial size and depth. In the next stage, where quantum feature encoding is performed, the features of the data to be classified are inserted as parameters of quantum gates, which perform various operations on these qubits. Therefore, a limited number of qubits also sets a bar on the total number of gate parameters; thus, the dimensionality reduction of classical data prior to its utilization within a quantum network is deemed imperative. 

Standard classical techniques, such as the autoencoder and simple resizing, are chosen as they allow for efficient compression of high-dimensional input data, which is important for reducing the computational complexity of quantum machine learning models. The autoencoder is particularly useful in this regard, as it can learn to represent the input data of dimensions $p\times p$ to a lower $q$-dimensional space, $q$ $<$ $p$, extracting a reduced set of features of size $q\times 1$, while still preserving important features and minimizing information loss. As an alternative, the simple resizing operation can also be effective in reducing the dimensionality of input data. It converts input data of dimensions $p\times p$ to a desired dimension of $q\times q$, $q$ $<$ $p$.
\subsubsection{Quantum Feature Encoding}
\begin{figure}[b]
    \centering
    \includegraphics[width=\linewidth]{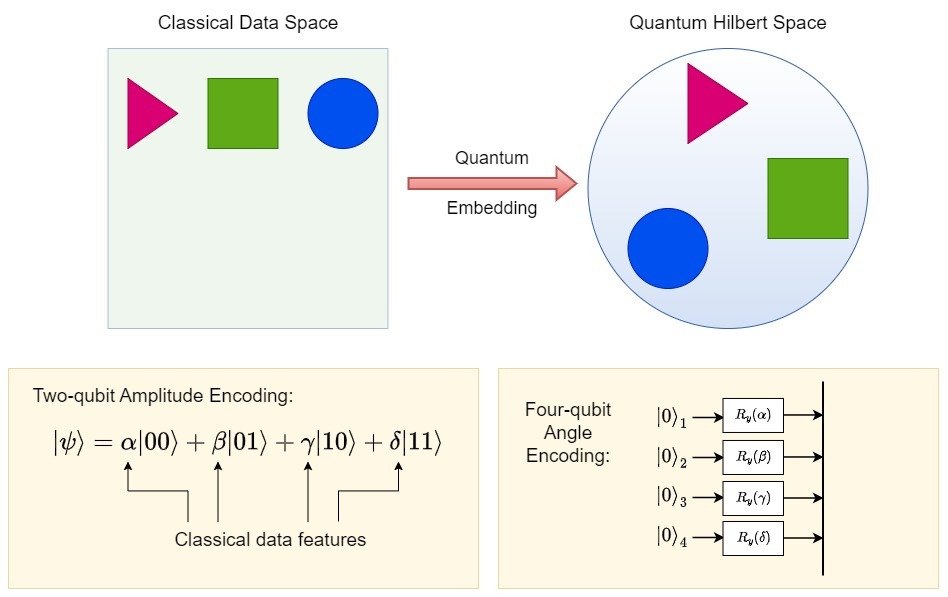}
    \caption{(Top) The Quantum Feature encoding for quantum machine learning maps classical data in the classical data space to quantum states in the Hilbert space. The different shapes denote different data points. On the right side, in the quantum Hilbert space, each shape represents the corresponding quantum composite state (Bottom left) \textit{Amplitude Encoding} is an example of such an encoding technique where $2^n$ (here n=2) data points can be mapped into n-qubits. (Bottom right) In contrast, \textit{Angle Encoding} maps the n-datapoints as arguments of $R_y$ gates to n qubits.}
    \label{fig:encoding}
\end{figure}
The projection of the reduced classical values, received as output from the previous layer, into quantum states is referred to as quantum feature encoding. Mathematically, the mapping of the classical input data, \textit{X}, into higher dimensional quantum states, represented in the Hilbert space and denoted by \textit{H}, is represented as \[\phi: X\mapsto H\]
where \(\phi\) is the feature map. In this stage of the network, $n$ qubits initialized to the state of\(\ \big|0\rangle\) are fed. 
The qubits are then subjected to state operations via quantum gates, parameterized by the classical data $X$, which is the output of the block performing classical data reduction. This results in the mapping of the classical data to the Hilbert space and the resulting state is represented by \(\big|\psi_x\rangle\), where $x$ is the classical data point from the image. This process of quantum state preparation encodes the classical values into the input qubits, which can then exploit the unique properties of superposition, entanglement, and interference. This process is visualized in Fig. \ref{fig:encoding}. There are several techniques and quantum ansatzes that are used to accomplish this task. A recent innovative work, \cite{schuld2021effect}, investigates the flexibility of quantum circuits to learn any function for a set of inputs in a framework focused on data encoding. Another paper, \cite{nguyen2022quantum}, proposes an automatic search algorithm to design the quantum network for embedding. However, two of the most common encodings, \textit{Amplitude} and \textit{Angle Encoding}, are selected for this work for the purpose of highlighting the performance of the proposed QCNN structure. Selecting the two encoding schemes further facilitates the comparability of the model as they are the most common encoding techniques adopted in QML literature. These encoding techniques are discussed in subsequent sections.


\paragraph{Amplitude Encoding}
In this encoding scheme, the normalized classical vectors from the Data Preprocessing Layer are represented as amplitudes of the \(n\)  input qubits in the Quantum Feature Encoding Layer. This displays a particular quantum advantage as normalized feature vectors of size \(2^n\) can be encoded into only $n$-qubits (\cite{schuld2018supervised}). The following equation shows the states prepared after performing \textit{Amplitude Encoding} on the input qubits. 
\begin{equation}
 \big|\psi_x\rangle\ = \sum_{i=1}^{N} x_i\big|i\rangle
\end{equation}
Here \(\big|\psi_x\rangle\) is the quantum state corresponding to the $N$-dimensional classical datapoint \(X\) after reduction, where \(N=2^n\), \(x_i\) is the \textit{i}-th element of the datapoint \(X\) and\(\ \big|i\rangle\) is the \textit{i}-th computational basis state. 

In a classical neural network, each binary value necessitates a distinct trainable weight or bias, resulting in a considerable number of parameters. In contrast, \textit{Amplitude Encoding} permits the representation of data through the amplitudes of a limited number of quantum states, thereby enabling a more compact representation (Fig. \ref{fig:encoding} bottom left). 

This has been demonstrated to result in a significant decrease in the number of trainable parameters, contributing to the simplification of the model and enhancement of its performance. While this method provides this benefit, it also increases the depth of the quantum circuit as \textit{O(poly(n))} or as \textit{O(n)} if the number of qubits fed in this layer is increased (\cite{div_conq}).

\paragraph{Angle Encoding}
\textit{Angle Encoding} is another technique employed in quantum machine learning for the representation of data, which utilizes the rotation of quantum gates ($R_x$, $R_y$ and $R_z$) to encode classical information (Fig. \ref{fig:encoding} bottom right). This method involves encoding the $N$ features of classical data as the angles of \(n\) input qubits between quantum states (\cite{schuld2021supervised}). In this method, $N$ has been kept equal to $n$ to allow us to use the maximum size of classical features possible. The advantage of this approach lies in its ability to represent continuous data more naturally and efficiently compared to \textit{Amplitude Encoding} (\cite{schuld2021supervised}). The states resulting from performing \textit{Angle Encoding} on the input qubits are:
\begin{equation}
 \big|\psi_x\rangle\ = \otimes_{i=1}^{n} R(x_i)\big|0^n\rangle  
\end{equation}
Here \(R(.)\) can be either of the rotation gates \(R_x\), \(R_y\), or \(R_z\).
In \textit{Angle Encoding}, the angles between the quantum states can be varied continuously to capture the intricacies of the data. This leads to a more precise and nuanced representation of the data and can result in improved performance for certain types of quantum machine learning models. Although, unlike \textit{Amplitude Encoding}, it can only encode one qubit with one feature value, resulting in the reduction of noise, which makes it particularly advantageous in NISQ computing. 

The selection of encoding techniques for this design is contingent upon the classical dimensionality reduction technique employed in the first layer. It can be recalled from the previous section that the \textit{Amplitude Encoding} method, which uses $n$ input qubits, can accommodate a maximum of $2^n$ data points. This requires the use of simple resizing to $2^{n/2} \times 2^{n/2}$ dimension followed by flattening, which is essential according to this state preparation method. Conversely, the \textit{Angle Encoding} technique encodes the flattened $N$ data points into $n$ qubits and thus relies on the use of an autoencoder to reduce the dimensions accordingly.

\subsection{The Convolutional Subsystem}

The family of QNNs that are tree-like in shape and rely on decreasing the number of qubits by a factor of 2 in each subsequent layer is known as Quantum Convolutional Neural Networks (QCNNs). This progressive reduction in qubits is similar to the pooling operation in classical CNNs. The conventional QCNN comprises only the quantum convolutional and pooling layers as the building blocks of such networks. As shown in Fig. \ref{fig:conv1}, the proposed model has a similar structure between the encoding layer and the classifier. Although the model with eight qubits is reduced to four using a pooling layer (Fig. \ref{fig:ansatz}), in the spirit of the conventional QCNN, the structure also contains the Interaction Layers (Fig. \ref{fig:entang_layer}) with extended qubit interactions.

The classical-data-modulated quantum states from the previous Encoding Subsystem flow into the convolutional and pooling layers sequence and are subjected to the ansatzes' unitary operations. The quantum state resulting from the convolutional or pooling layer is expressed as:
\begin{equation} 
  \left|\psi(\theta_i)\right\rangle \left\langle\psi(\theta_i)\right| = \mathrm{Tr}_{A_i} \left( U_{\theta_{i-1}} \left|\psi(\theta_{i-1})\right\rangle \left\langle\psi(\theta_{i-1})\right| U_{\theta_{i-1}}^\dagger \right)
\end{equation}
Where $\big|\psi_{i-1}\rangle$ is the input state, $\big|\psi_i\rangle$ is the output state of the layer, $U(\theta)$ is the parameterized unitary operation of the layer, and $\mathrm{Tr}_{A_i}(\cdot)$ is the partial trace operation over subsystem $A_i$. This derives the reduced state of the system, excluding any desired subsystem denoted by  $A_i$. It must be noted that $\big|\psi\rangle$ is the composite space of $n$ qubits involved in the system, $\big|\psi\rangle^{\otimes n}$. A complete representation of all (parameterized) gates $U(\theta)$ is summarized in table: \ref{tab:A}, appendix A.

\subsubsection{The Quantum Convolutional Layer}
\begin{figure}[b]
    \centering
    \includegraphics[width=\linewidth]{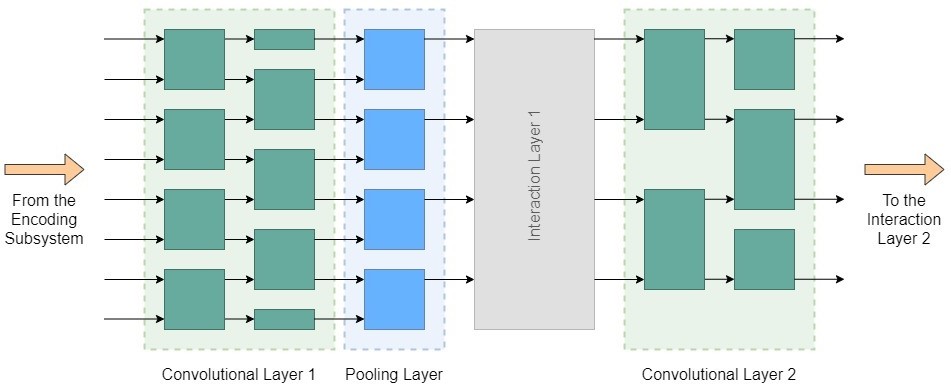}
    \caption{The Convolutional Layers comprises ansatzes, all of which use two-qubit interactions. Each green box represents either convolutional ansatz 1 or 2. Note that in the second vertical line of the Convolutional Layers, the topmost and bottom green boxes denote the two halves of the same ansatz, which means that the first and last qubits are fed as input to the same ansatz. The blue boxes in the Pooling Layer are made up of the pooling ansatz, and they reduce the number of qubits.}
    \label{fig:conv1}
\end{figure}
\begin{figure}[t]
    \centering
    \[ \Qcircuit @C=0.7em @R=1.1em { 
    & \gate{U_3(\theta_1,\theta_2,\theta_3)} & \ctrl{1} & \gate{R_y(\theta_7)} & \targ & \gate{R_y(\theta_9)} & \ctrl{1} & \gate{U_3(\theta_1{}_0,\theta_1{}_1,\theta_1{}_2)} & \qw \\
    & \gate{U_3(\theta_4,\theta_5,\theta_6)} & \targ & \gate{R_z(\theta_8)} & \ctrl{-1} & \qw & \targ & \gate{U_3(\theta_1{}_3,\theta_1{}_4,\theta_1{}_5)} & \qw
    } \]

    \centering
    \[ \Qcircuit @C=1em @R=1.1em {
     & \gate{R_x(\theta_1)} & \gate{R_z(\theta_2)} & \gate{R_x(\theta_3)} & \ctrl{1} & \gate{R_x(\theta_4)} & \gate{R_z(\theta_5)} & \qw \\
     & \gate{R_x(\theta_6)} & \gate{R_z(\theta_7)} & \ctrl{-1} &  \gate{R_x(\theta_8)} & \gate{R_x(\theta_9)} & \gate{R_z(\theta_1{}_0)} & \qw
     } \]

    \centering
    \[ \Qcircuit @C=1.2em @R=2.1em {
    \lstick{\ket{\psi_1'}} & \qw & \gate{R_z(\theta_1)} & \qw & \qw & \qw & \gate{R_x(\theta_2)} & \rstick{\ket{\psi_1''}} \qw\\
     \lstick{\ket{\psi_n'}} & \qw & \ctrl{-1} & \qw & \gate{X} & \qw & \ctrlo{-1} & \qw \qw
    } \]
    \caption{(Top) Ansatz 1 contains 15 trainable parameters. (Middle) Ansatz 2 contains 10 trainable parameters. These two ansatzes are both used for building the Convolutional Layers. (Bottom) The ansatz used for building the Pooling Layer. The gates used to construct the ansatzes are summarized in table \ref{tab:A}.}
    \label{fig:ansatz}
\end{figure}
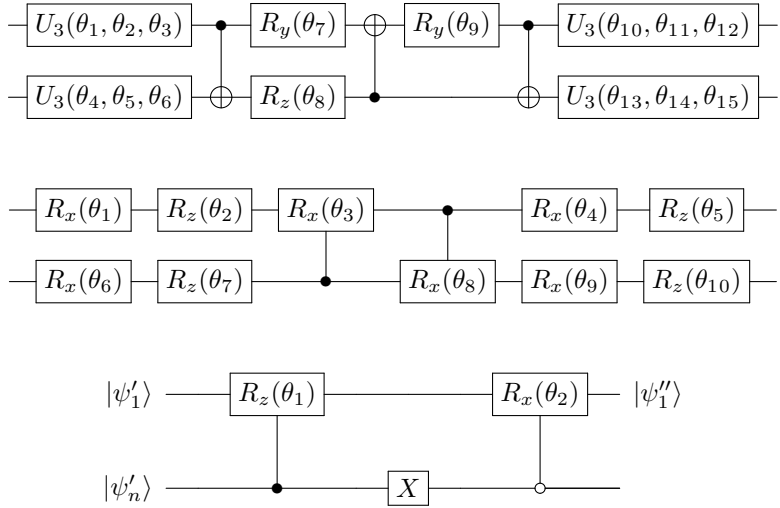
Convolutional Layers 1 and 2 are connected to the output of the Encoding Layer and the first Interaction Layer, respectively, and have an internal structure as shown in the simplified block diagram Fig. \ref{fig:conv1}. The building blocks for a particular Convolutional Layer are called parameterized quantum circuits called ansatzes, which, for any instance, are kept the same for both the Convolutional Layers. The ansatzes used to construct these layers are shown in Fig. \ref{fig:ansatz}. As shown in Fig. \ref{fig:conv1}, each Convolutional Layer has multiple identical convolutional ansatz blocks, which act like two-qubit gates and are applied to the nearest neighbor qubits in a translationally invariant way. This means that the first vertical line of ansatzes connects each qubit to one of its nearest neighbors. Then, using identical blocks again, the second vertical line of ansatzes is cascaded with the first to connect each qubit to its other nearest neighbor. These two lines of ansatzes causing complete connection of each qubit to their nearest neighbors result in translation invariance, an advantageous trait of conventional QCNN, and is defined in this work as a Convolutional Layer. It must be noted that all the ansatz blocks are identical and share the exact weights in each Convolutional Layer.

The first ansatz in Fig. \ref{fig:ansatz} consists of a relatively large number of parameters, i.e., 15, which helps increase flexibility, and the controlled \textit{R} gates help increase expressibility. The second ansatz has five fewer parameters, which is a parametrized form of a reduced version of the circuit that recorded the best expressibility in a study carried out by \cite{sim2019expressibility}. The ansatzes consist of the $R_x$, $R_y$, $R_z$ gates, which cause qubit rotations about the $x$, $y$, and $z$ axes, respectively. Ansatz 1 additionally has the $U3$ gate ($3$ representing the number of parameters), which can be decomposed to rotation and phase shift gates and is represented by the matrix with respect to the computational basis as follows:
\begin{equation}
 U{3}(\theta_1,\theta_2,\theta_3) = \begin{bmatrix} \cos(\frac{\theta_1}{2}) & -\text{e}^{i\theta_3}\sin(\frac{\theta_1}{2}) \ \\
\text{e}^{i\theta_2}\sin(\frac{\theta_1}{2}) & \text{e}^{i(\theta_2+\theta_3)}\cos(\frac{\theta_1}{2}) \end{bmatrix}   
\end{equation}

Here, $\theta_1$, $\theta_2$, and $\theta_3$ are the parameters of the $U3$ gate, and the matrix represents a unitary operation on a qubit. The selection of two different ansatzes is to inspect the flexibility of the proposed network performance on slight changes in the structure of the ansatz and the number of trainable parameters. 


\subsubsection{The Quantum Pooling Layer}
The Quantum Pooling Layer is placed after the first Quantum Convolutional Layer to reduce the number of qubits to half (Fig. \ref{fig:conv1}). The primary purpose of the Pooling Layer in any convolutional neural network is to reduce the data representation's spatial size and maintain the most critical information. In the process, the layer helps reduce the computational cost of the network and improve its generalization capabilities by decreasing overfitting, making it robust to translations, rotations, and other minute changes. 

The quantum Pooling Layer in Fig. \ref{fig:ansatz} (bottom) traces out one qubit from the two qubits it is fed and thus reduces the two-qubit states to a one-qubit state. The Pooling Layer uses two controlled rotation gates and a $Pauli-X$ gate. The Pooling Layers and the Convolutional Layers extend qubit interactions beyond nearest neighbors, establishing further dependencies. Just like the Convolutional Layer, the Pooling Layer ansatzes (denoted by blue boxes in Fig. \ref{fig:conv1}) share the exact same weights. An Interaction Layer further processes the output qubits to aid the learning process.

\subsubsection{The Interaction Layers}
\begin{figure}[t]
    \centering
    \includegraphics[width=\linewidth]{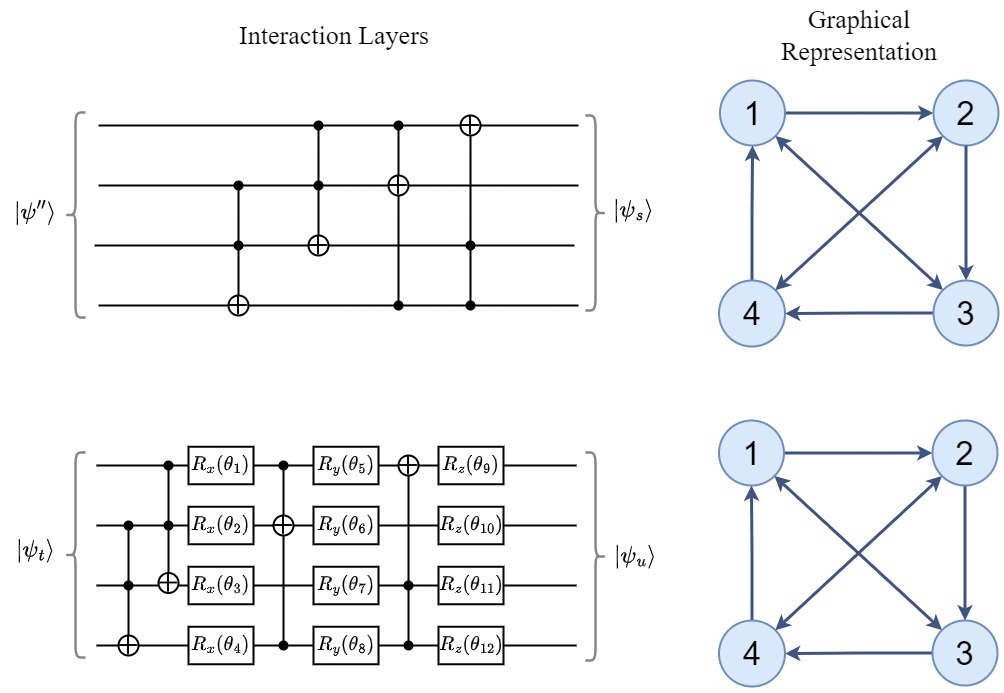}
    \caption{The proposed Interaction Layer 1 (at the top), which comes after the first Convolutional Layer and is followed by the second Convolutional Layer. The novel Interaction Layer 2 (at the bottom) comes after the second Convolutional Layer and is followed by the final Classifier Subsystem. The graphical representations show the dependency of each state established via Toffoli gates in each layer. Each state depends on two target states. For e.g., a Pauli-X operation will be carried on qubit 4 if both of the 2nd and 3rd qubits are in a state \ket{1}. }
    \label{fig:entang_layer}
  \end{figure}

The novelty brought forward in the proposed quantum architecture involves making variational quantum layers designed to introduce extensive entanglement and expressibility in the overall quantum network.
In order to bring forward this unique quantum phenomenon, in this proposed \textit{Interaction layer}, \textit{Toffoli} gates are cascaded with the convolutional and rotational gates. They are expected to establish three-qubit interactions, as shown in Fig.  \ref{fig:entang_layer}.


In this era of NISQ computing, implementing two-qubit gates is more complex than single-qubit ones. However, with the recent developments in quantum hardware and the promise of near error-free quantum computers in the not-so-distant future, using multi-qubit gates in quantum computers is expected to be more common. It must be noted that the use of two-qubit gates, such as the \textit{CNOT} gate, in quantum machine learning models is motivated by the increase in entanglement and expressibility of the network, which helps it to learn more complex features of classical data \cite{sim2019expressibility}. Therefore, studies must be conducted exploring the effectiveness of three-qubit gates in various quantum networks. 

This paper explores the comparative advantage of three-qubit \textit{Toffoli} gates by introducing Interaction layers in the conventional QCNN structure. The difference in performance upon this addition is expected to indicate the extent of effectiveness that may result from the successful implementation of such gates in quantum hardware.  

In the first Interaction Layer, the four \textit{Toffoli} gates entangle the four qubits in a circuit-block interaction configuration, as illustrated in Fig. \ref{fig:entang_layer}, which means that interdependency has been established between these quantum states, so the measured value of one state will depend on the others. This particular configuration is chosen over a nearest-neighbor or all-to-all configuration, as it displays favorable expressibility and less qubit connectivity requirements. The placement of these \textit{Toffoli} gates after the previous layers enables the network to span all the basis states more strongly (i.e., with a higher probability for the basis states that previously had near-zero probabilities) than without them. The output state $\ket{\psi''}$ of the pooling layer is input to Interaction Layer 1. The output state of this Interaction layer $\ket{\psi_s}$ is derived from the input state $\ket{\psi''}$ by the following equation.
\begin{equation}
    \ket{\psi_s} = U_{t341} U_{t123} U_{t142} U_{t234} \ket{\psi''} \label{5}
\end{equation}
Where $U_{txyz}$ represents the corresponding Unitary matrix to the \textit{Toffoli} gate (\ref{secA1}) and $x,y,z$ are the control 1, control 2, and target qubit number in the composite space, respectively. The $\ket{\psi''}$ is the input to, and $\ket{\psi_s}$ is the output of the Interaction Layer 1. The working principle of the \textit{Toffoli} gate is shown as the graphical representation in Fig. \ref{fig:entang_layer2}.

Indeed, the \ket{\psi_s} is still a $4$ qubit composite state
\begin{equation}
    \ket{\psi_s} = \sum_{e_1, e_2, e_3, e_4} c_{e_1 e_2 e_3 e_4} \ket{e_1 e_2 e_3 e_4}
\end{equation}
Here, the $e_1, e_2, e_3, e_4$ represent the basis vectors for the 4 qubits.


After the second Convolutional Layer, the qubit state $\ket{\psi_t}$ is passed through Interaction Layer 2. This Interaction Layer differs from the first in the inclusion of \(R_x\), \(R_y\), \(R_z\) rotation gates with trainable parameters between its \textit{Toffoli} gates as shown in Fig. \ref{fig:entang_layer}. The parameterized gates between the \textit{Toffoli} gates increase the degree of freedom of the quantum states, increasing the flexibility of the learning process and, therefore, has the potential to learn more nuanced features of the training data. The particular difference between this layer and the Interaction Layer 1 can be seen from the following equation.
\begin{equation}
    \ket{\psi_u} = R_{z(\theta_9, \theta_{10}, \theta_{11}, \theta_{12})} U_{t341} R_{y(\theta_5, \theta_6, \theta_7, \theta_{8})} U_{t123} R_{x(\theta_1, \theta_2, \theta_3, \theta_{4})} U_{t142} U_{t234} \ket{\psi_t} \label{6}
\end{equation}
Where the $R_x, R_y, R_z$ gates are the rotational gates with parameters $\theta_w, \theta_x, \theta_y, \theta_z$ on qubits $1, 2, 3, 4$ respectively, $U_{txyz}$ represents the corresponding Unitary matrix to the \textit{Toffoli} gate ($x,y,z$ are the control 1, control 2, and target qubit number in the composite space). The $\ket{\psi_t}$ is the input to, and $\ket{\psi_u}$ is the output of the Interaction Layer 2. It can be noticed that it is these parameterized gates that differ eqn. \eqref{6} from eqn. \eqref{5}. A complete summary of the unitary gates is provided in table \ref{tab:A}.

\subsection{The Classifier Subsystem}
The last Interaction Layer in Fig. \ref{fig:entang_layer2} acts as a classifier, taking input of the modulated qubit state $\ket{\psi_u}$, (output of Interaction Layer 2), from the Convolutional Subsystem and tracing the output to three qubits. A single instance of the ansatz is shown in Fig. \ref{fig:entang_layer2}. The layer further has three rotational gates $R_x, R_y, R_z$ per ancilla qubit. 

\subsubsection{The Third Interaction Layer}
\begin{figure}[t]
    \centering
    \includegraphics[width=\linewidth]{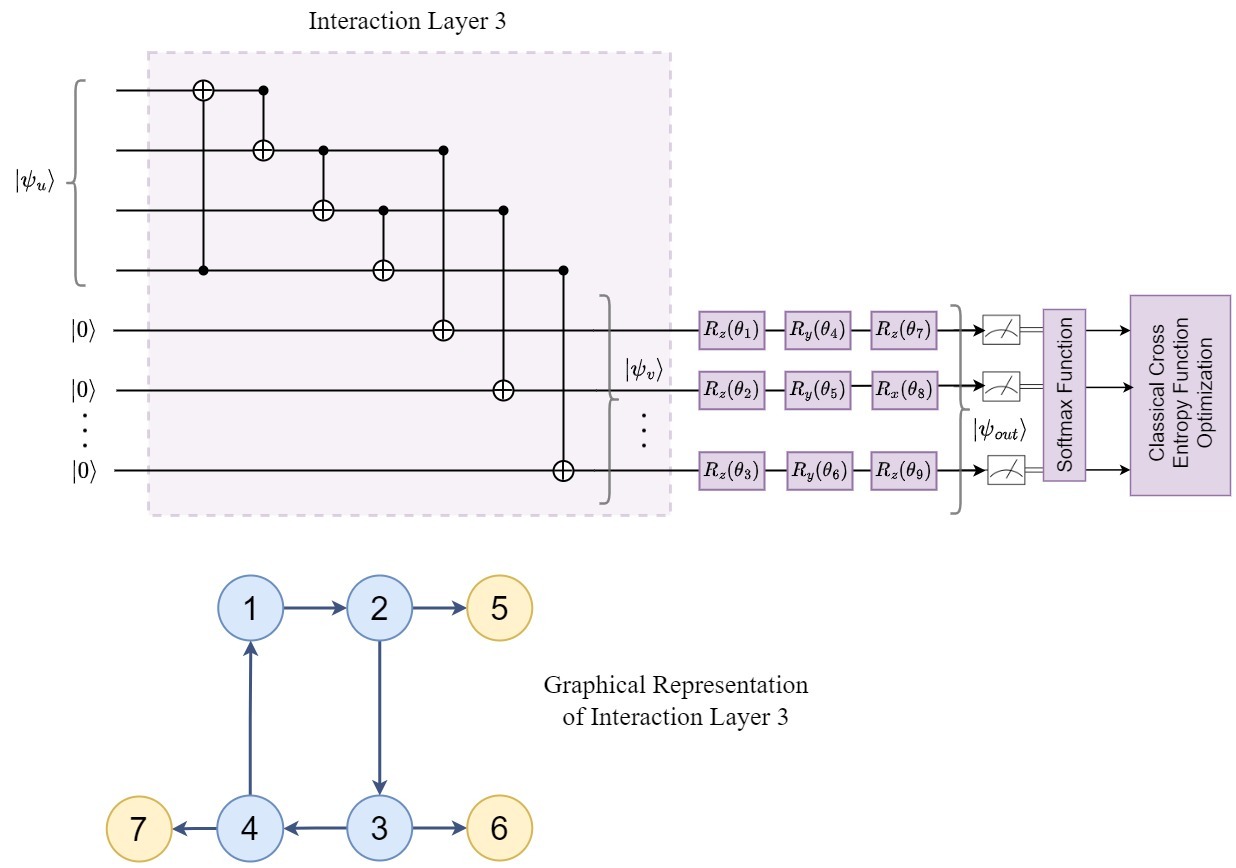}
    \caption{(Top) The Classifier Subsystem is designed such that it can classify two or more classes. It comprises the Interaction Layer 3, followed by rotation gates before being measured. Note that the qubits initialized to \ket{0} at the bottom are ancilla qubits where the number of ancilla qubits used equals the number of classes. (Bottom) The graphical representation of the entangling structure of the third Interaction Layer.}
    \label{fig:entang_layer2}
  \end{figure}

The third Interaction Layer utilizes \textit{CNOT} gates (table \ref{tab:A}) to entangle the remaining qubits with the ancilla qubits, which are used to store the entangled states. It can be observed that the number of ancilla qubits is equal to the number of classes that are to be classified using the network and have been set to \ket{0} initially. The ancilla qubits interact with the remaining qubits of the network through the \textit{CNOT} gates as shown in Fig. \ref{fig:entang_layer2} and are passed through the three rotational gates at the terminal of the quantum network. The rotational gates comprise $R_x, R_y, R_z$ gates with trainable parameters, further helping the flexibility of the training process. It must be noted that the new composite space has increased after the addition of ancilla qubits.
\begin{equation}
    \ket{\psi_v} = U_{c47} U_{c36} U_{c25} U_{c34} U_{c23} U_{c12} U_{c41} (\ket{\psi_u} \otimes \ket{\psi_{ancilla}})
\end{equation}
where \ket{\psi_v} is the output state of Interaction Layer 3, \ket{\psi_u} is the input state of Interaction Layer 3 and \ket{\psi_{ancilla}} is the initialized ancilla qubits. $U_{cxy}$ denotes CNOT gates with control on x qubit and target at y qubit. The tensor product between \ket{\psi_u} and \ket{\psi_{ancilla}} converts the composite space to higher dimensions, including the ancilla qubits, and consequently, the state is subjected to a series of \textit{CNOT} gates. Then, the new state is passed through rotational gates $R_x$, $R_y$, $R_z$. (table: \ref{tab:A})
\begin{equation}
    \ket{\psi_{out}} =  R_{x123} R_{y123} R_{z123} (\ket{\psi_v})
\end{equation}
Here, $\ket{\psi_{out}}$ is the output of the Classifier subsystem, and (\ket{\psi_v}) is the input to the rotational gates coming from Interaction Layer 3. The  $R_{x123} R_{y123} R_{z123}$ gates denote the equivalent rotational unitaries about the x, y, and z axes acting on all the ancilla qubits. Although the number of ancilla qubits shown here is three for simplicity, it should be equal to the number of classes to be classified.

\subsubsection{Measurement and Data Aggregation Method}

Finally, as the first step of the data aggregation method from quantum to classical values, the ancilla qubit states are measured as expectation values on the Pauli-Z operator, which causes the collapse of the quantum states to deterministic values. The choice of measuring the expectation value of the Pauli-Z operator is common in QML literature, as it corresponds to a measurement on the computational basis and provides a convenient way to calculate the probabilities of the basis states. The Pauli-Z operator has eigenvalues of +1 \& -1 corresponding to eigenvectors $\ket{0}$ \& $\ket{1}$, respectively. If the ancilla qubit to be measured has a state $|\psi\rangle$, 
then, the expectation value of the Pauli-Z operator can be calculated from the expression
\begin{equation}
    \langle Z \rangle = \langle \psi | Z | \psi \rangle
\end{equation}
where $Z$ is the Hermitian matrix, corresponding to the linear operator in the computational basis. The expectation value depends on the eigenvalues of the operator (-1 \& +1) and on the probability amplitudes of the measured ancilla qubit, resulting in a real value in range $\langle Z \rangle \in [-1, 1]$. This means that for $i$ number of classes, there are $i$ ancilla qubits, and therefore, the output of the network is an $i$ dimensional vector.



Following the measurement of the ancilla qubits, the classical values are sent to the \textit{softmax} function to calculate the probability vectors for each class.
\begin{equation}
    y_r = \text{softmax}(\mathbf{x})_r = \frac{e^{x_r}}{\sum_{j=1}^{i} e^{x_j}}
\end{equation}

\begin{equation}
    \overline{\mathbf{y}} = \begin{bmatrix} y_1 \\ y_2 \\ \vdots \\ y_i \end{bmatrix}
\end{equation}

The output of the Softmax function transforms the vector elements into a range $\text{softmax}(\mathbf{x})_r \in [0, 1]$. This normalizes the $i$ dimensional output vector, making it comparable to the labeled class of the one-hot-encoded input data. The output vector at this point is ready to be input to the cost function.

The cost function used specifically for this work is the classical categorical cross-entropy loss function, which can be expressed as:
\begin{equation}
 \text{loss} = \sum_{j=1}^{i} y_j.log(\bar{y_j})
\end{equation}
where \(y_j\) is an element of the one-hot-encoded ground truth vector $\mathbf{y}$ of dimension $i$, \(\bar{y_j}\) is the predicted probability of the corresponding class and $i$ is the total number of classes. The parameters of the quantum gates are optimized through gradient descent using classical computational techniques, after which the parameters are updated accordingly through back-propagation.

The introduction of these Interaction Layers in the middle of the conventional QCNN, along with an ancilla-based classifier (Interaction Layer 3), is expected to provide promising results. It is inspected in detail later in the next section \ref{sec:sim_results}. The proposed Interaction Layers consisting of a combination of \textit{Toffoli} and \textit{CNOT} gates along with trainable parameters in between the Convolutional Layers, and the use of ancilla qubit-\textit{CNOT} classifier in QCNNs is a first to the best of our knowledge. The architecture is summarized in table: \ref{tab:summary}.

\begin{table}[ht]
\caption{Tabular representation of the architectural workflow, with a focus on dimensions and parameters tailored for the Binary Classification task. The operations are arranged in a sequence, starting from the top and progressing to the bottom. In the Operation Sequence column, additional information regarding the classical data dimension or the type of encoding/ansatz is mentioned. The only difference in parameters for Multiclass classification is the addition of 3 parameters per new class in Interaction Layer 3.}
    \centering
    \renewcommand{\arraystretch}{1.75} 
    \begin{tabular}{|c|c|c|}
        \hline
        \textbf{Operation Sequence} & \textbf{\begin{tabular}[c]{@{}c@{}} Processing Size of \\ the Quantum \\ Composite State\end{tabular}} & \textbf{\begin{tabular}[c]{@{}c@{}}Trainable \\ Parameters\end{tabular}} \\
        \hline
        Classical Data ($28 \times 28$) & - & - \\
        \hline
        \begin{tabular}[c]{@{}c@{}}Data Preprocessing \\ (Autoencoder/Resize $1\times8$ / $16\times16$)\end{tabular} & - & 0 \\
        \hline
        Encoding (Amplitude/Angle) & 8 qubits & 0 \\
        \hline
        Quantum Conv. Layer 1 (ansatz 1/2) & 8 qubits & 15/10 \\
        \hline
        Pooling & 4 qubits & 2 \\
        \hline
        Interaction Layer 1 & 4 qubits & 0 \\
        \hline
        Quantum Conv. Layer 2 & 4 qubits & 15/10 \\
        \hline
        Interaction Layer 2 & 4 qubits & 12 \\
        \hline
        \begin{tabular}[c]{@{}c@{}}Interaction Layer 3\\ (previous layer output + ancilla qubits)\end{tabular} & 4 qubits + 2 qubits & 6 \\
        \hline
        Measurement & 2 qubits & 0 \\
        \hline
        SoftMax ($2\times1$)& - & - \\
        \hline
        \textbf{Total Parameters} \textbf{(ansatz 1/2)} & - & \textbf{50/40} \\
        \hline
    \end{tabular}
    \label{tab:summary}
\end{table}

\section{Simulation and Results}\label{sec:sim_results}

\subsection{Dataset}
The widely utilized standard datasets, namely \textit{MNIST} \cite{deng2012mnist} and \textit{Fashion MNIST} \cite{xiao2017fashion} are employed to benchmark the proposed QCNN model. \textit{MNIST} stands for Modified National Institute of Standards and Technology and has been developed to be used as a benchmarking dataset for various machine learning models. Binary classification involving classes (0, 1) and three-class classification involving classes (0, 1, 2) are performed. \textit{MNIST dataset} consists of grayscale images of handwritten digits for ten classes from 0 to 9. The number of training (test) images for class 0 is 5,923 (980); for class 1, it is 6,742 (1,135); for class 2, it is 5,958 (1,032). The \textit{Fashion MNIST} dataset consists of images of ten classes of clothing items. In the relevant classes for this work, class 0 is a shirt/top, class 1 is a trouser, and class 2 is a pullover. It is made up of 6,000 training and 1,000 test images per class. The original size of the images from either dataset is $28\times28$, and a reduction in dimension is accomplished through a classical autoencoder or simple resize to the desired shape. A third dataset known as the \textit{Iris dataset} \cite{de2015mobile} is used solely for the purpose of multiclass classification. It consists of feature data of three classes of iris species with 50 samples per class. The features include 4 attributes per sample, namely sepal length, sepal width, petal length \& petal width. The dataset is such that one flower class is linearly separable from others, but the other two classes are not linearly separable from each other.

\begin{table*}[]
\caption{Table showing the details of the datasets used and the classical data preprocessing methods employed upon them. Note that no data preprocessing is needed for \textit{Iris dataset} as is directly input in the \textit{Angle Encoding} block.}
\centering
\resizebox{\columnwidth}{!}{%
\begin{tabular}{cccccccc}
\hline
\multirow{2}{*}{\textbf{Type}} &
  \multirow{2}{*}{\textbf{Name}} &
  \multirow{2}{*}{\textbf{Size}} &
  \multirow{2}{*}{\textbf{Classes used}} &
  \multicolumn{2}{c}{\textbf{Sample Size}} &
  \multicolumn{2}{c}{\textbf{Preprocessing}} \\ 
 &
   &
   &
   &
  \textbf{Training} &
  \textbf{Testing} &
  \textbf{Method} &
  \textbf{\begin{tabular}[c]{@{}c@{}}Size of output\\ features\end{tabular}} \\ \hline\\
\multirow{4}{*}{Image} &
  \multirow{2}{*}{\makecell{Fashion\\ MNIST}} &
  \multirow{4}{*}{$28\times28$} &
  \multirow{4}{*}{\begin{tabular}[c]{@{}c@{}}binary: 0 \& 1\\ multi: 0, 1 \& 2\end{tabular}} &
  \multirow{2}{*}{6000/class} &
  \multirow{2}{*}{1000/class} &
  Resize &
  $16\times16$ \\ 
 &
   &
   &
   &
   &
   &
  Autoencoder &
  $8\times1$ \\  \\
 &
  \multirow{2}{*}{MNIST} &
   &
   &
  \multirow{2}{*}{\begin{tabular}[c]{@{}c@{}}0: 5923\\ 1: 6742\\ 2: 5958\end{tabular}} &
  \multirow{2}{*}{\begin{tabular}[c]{@{}c@{}}0: 980\\ 1: 1135\\ 2: 1032\end{tabular}} &
  Resize &
  $16\times16$ \\ 
 &
   &
   &
   &
   &
   &
  Autoencoder &
  $8\times1$  \\\\\\ \hline\\
Feature &
  Iris &
  $4\times1$ &
  multi: 0, 1 \& 2 &
  113 &
  37 &
  - &
  - \\ \\\hline
\end{tabular}%
}
\label{tab:dataset}
\end{table*}


\subsection{Simulation}
The simulation of the proposed QCNN model is conducted using the Pennylane library(\cite{bergholm2018pennylane}) version 0.28.0 and written in Python 3.7.12. The quantum simulator provided by Pennylane, known as the default$.$qubit, is used for the simulation process. The variational circuit is trained through the use of the Nesterov Moment Optimization algorithm (\cite{nesterov}). Initially, the labels of the classes are converted to one hot label vector of length equal to the number of classes and are treated as ground truth values when calculating the cost. A loop is then executed through the training process where a batch of randomly selected images is fed into the network in each iteration, reducing run time and preventing the gradient from becoming trapped in a local minimum. The optimization of the learning process is further facilitated through the use of an adaptive learning rate, where the learning rate is decreased as the rate of change of the output of the cost function is decreased. This training process is run for 1000 iterations, and test accuracy on the test images is calculated every 10 iterations. The mean test accuracy for each circuit and dataset configuration reported in the next section is based on an average of 10 independent runs with different random initializations. Further details regarding the batch size and learning rate of the optimizer vary for different cases and are therefore mentioned in the subsequent relevant sections.

\subsection{Performance evaluation}
\subsubsection{Binary classification}
\begin{figure}[t]
\centering
\includegraphics[width=\linewidth]{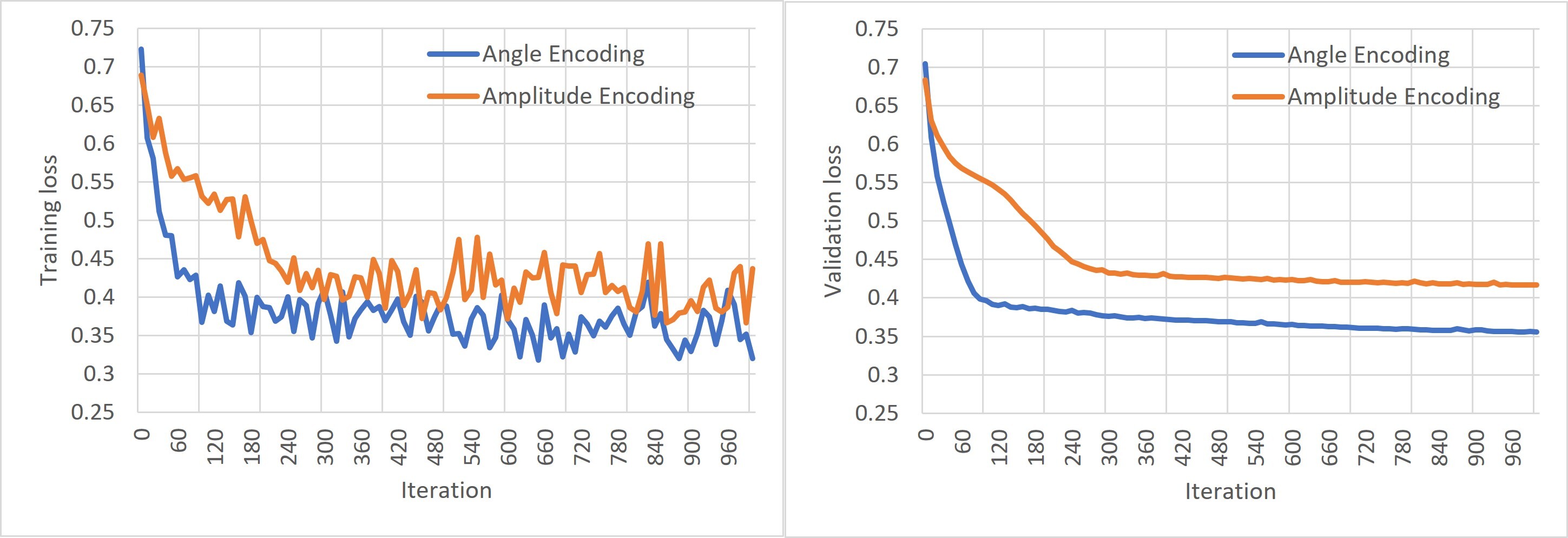}
\caption{Plots showing the loss for the binary classification of training images (left) and testing images (right) of the \textit{Fashion MNIST} dataset over 1000 iterations using Ansatz 1. Both show that for \textit{Amplitude Encoding}, the costs converge to a lower value. The convergence also occurs more rapidly when \textit{Amplitude Encoding} is used for the \textit{Fashion MNIST} dataset.}
\label{fig:ans1_losses}
\end{figure}
\begin{figure}[!ht]
\centering
\includegraphics[width=\linewidth]{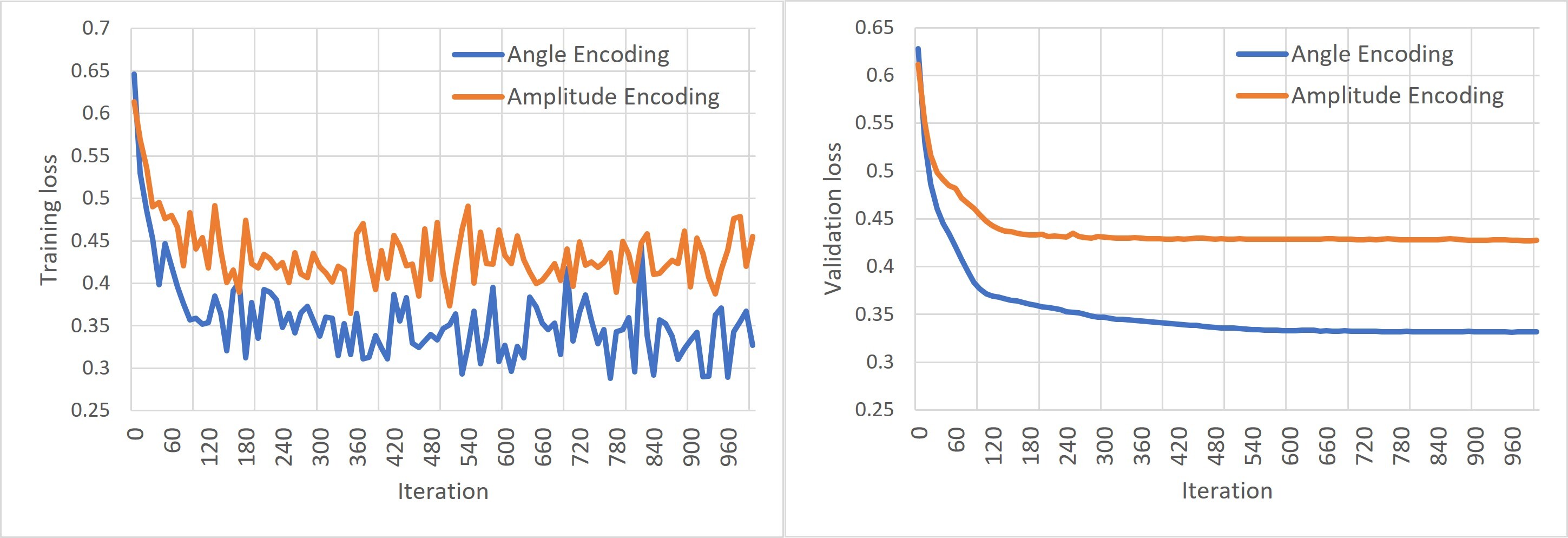}
\caption{Plots showing the loss for the binary classification of training images (left) and testing images (right) of the \textit{Fashion MNIST} dataset over 1000 iterations using Ansatz 2. Both show that for \textit{Amplitude Encoding}, the costs converge to a lower value. The convergence also occurs more rapidly when \textit{Amplitude Encoding} is used for the \textit{Fashion MNIST} dataset.}
\label{fig:ans2_losses}
\end{figure}
For the binary classification problem, classes 0 and 1 are chosen for both datasets in order to compare them with previous works. A total of eight input qubits, along with two ancilla qubits for the two classes, are used in the proposed model. The Convolutional and the Pooling Layers are arranged as illustrated in Fig. \ref{fig:overall_network}. 
During training, the batch size is kept at 50 images, which are randomly selected in each iteration. The learning rate in the Nesterov Optimizer is tuned to be 0.05 at the beginning of the learning process. In the later stages, it is reduced in accordance with the decrease in the cost and the improvement of test accuracy. Specifically, it is halved after 50 iterations and further halved after 100 iterations, after which it is kept constant. The trainable parameters are initialized randomly using the normal distribution, and the average classification accuracy is calculated over $10$ random initializations.
\begin{table*}[t]
    \caption{Results of the proposed binary classification model applied on two datasets. The mean accuracy of classifying test data and the standard deviation have been calculated over 10 independent runs.}
    \centering
    \resizebox{\columnwidth}{!}{%
    \begin{tabular}{cccccc}
    \hline
    \textbf{Dataset}               & \textbf{Ansatz}    & \makecell{\textbf{Data}\\\textbf{Preprocessing}} & \makecell{\textbf{Encoding}\\\textbf{Method}} & \makecell{\textbf{Quantum Gate}\\\textbf{Parameters}} & \makecell{\textbf{Text}\\\textbf{Accuracy(\%)}} \\ \hline
    \multirow{4}{*}{Fashion MNIST} & \multirow{2}{*}{1}& Autoencoder & Angle & \multirow{2}{*}{50}      & $95.75\pm{0.80}$                  \\ 
     &                    & Resize & Amplitude &                     & $92.89\pm{0.54}$ \\ 
     & \multirow{2}{*}{2} & Autoencoder & Angle       & \multirow{2}{*}{40} & $94.11\pm{0.64}$ \\  
     &                    & Resize & Amplitude &                     & $91.80\pm{0.71}$ \\ \hline
    \multirow{4}{*}{MNIST}         & \multirow{2}{*}{1} & Autoencoder & Angle  & \multirow{2}{*}{50}      & $98.16\pm{0.65}$                 \\ 
     &                   &Resize & Amplitude &                     & $99.00\pm{0.91}$  \\  
     & \multirow{2}{*}{2} &Autoencoder & Angle       & \multirow{2}{*}{40} &  $96.84\pm{1.50}$    \\ 
     &                   &Resize & Amplitude &                     &    $94.87\pm{0.81}$   \\ \hline
    \end{tabular}%
    }
    \label{tab:1}
\end{table*}

\begin{figure}[t]
\centering
\includegraphics[scale = 0.7]{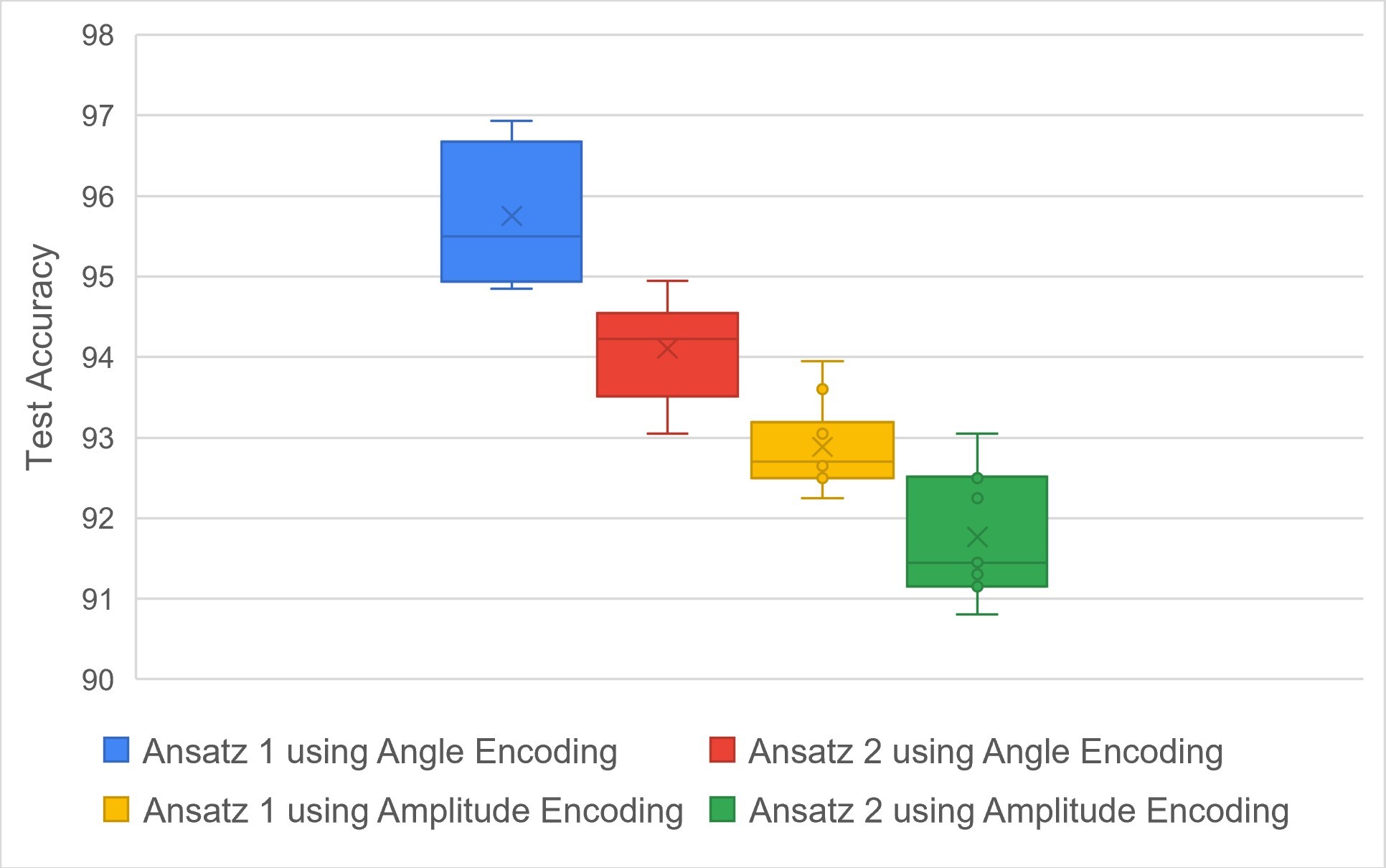}
\caption{Accuracy results using different encoding techniques and different ansatzes for binary classification on the test images of the \textit{Fashion MNIST} dataset. It shows that higher accuracies are obtained using \textit{Angle Encoding} over \textit{Amplitude Encoding} for the \textit{Fashion MNIST} dataset. Also, in both cases of encoding, Ansatz 1 results in higher accuracy due to its extra parameters. The results are computed over 10 independent runs.}
\label{fig:acc_angvsamp}
\end{figure}

The effect of the following different approaches on overall performance is investigated:

1. Quantum encoding by either \textit{Amplitude Encoding} or \textit{Angle Encoding}.

2. The two parameterized ansatzes given in Fig. \ref{fig:ansatz} used to construct the Convolutional Layers in Fig. \ref{fig:conv1}. 

  
The classification of the \textit{Fashion MNIST} is benchmarked to an accuracy of $95.75\%\pm0.80\%$ using the combination of autoencoder with \textit{Angle Encoding} and ansatz 1 as the Convolutional Layer filters, for which the total number of trainable parameters in the quantum network is 50. The training and validation losses are graphically shown in Fig. \ref{fig:ans1_losses} and Fig. \ref{fig:ans2_losses} using ansatz 1 and ansatz 2, respectively, for both \textit{Angle} and \textit{Amplitude Encoding}. It can be seen that \textit{Angle Encoding} performs better independent of ansatzes for the \textit{Fashion MNIST dataset}. On the other hand, peak accuracy attained in Binary classification for the \textit{MNIST} dataset is $99.00\%\pm0.91\%$. The number of trainable parameters, in this case, is also 50, and simple resizing + \textit{Amplitude Encoding} with ansatz 1 is used. 

The accuracies for the different combinations are summarized in Table \ref{tab:1}.
\begin{figure}[t]
\centering
\includegraphics[width=\linewidth]{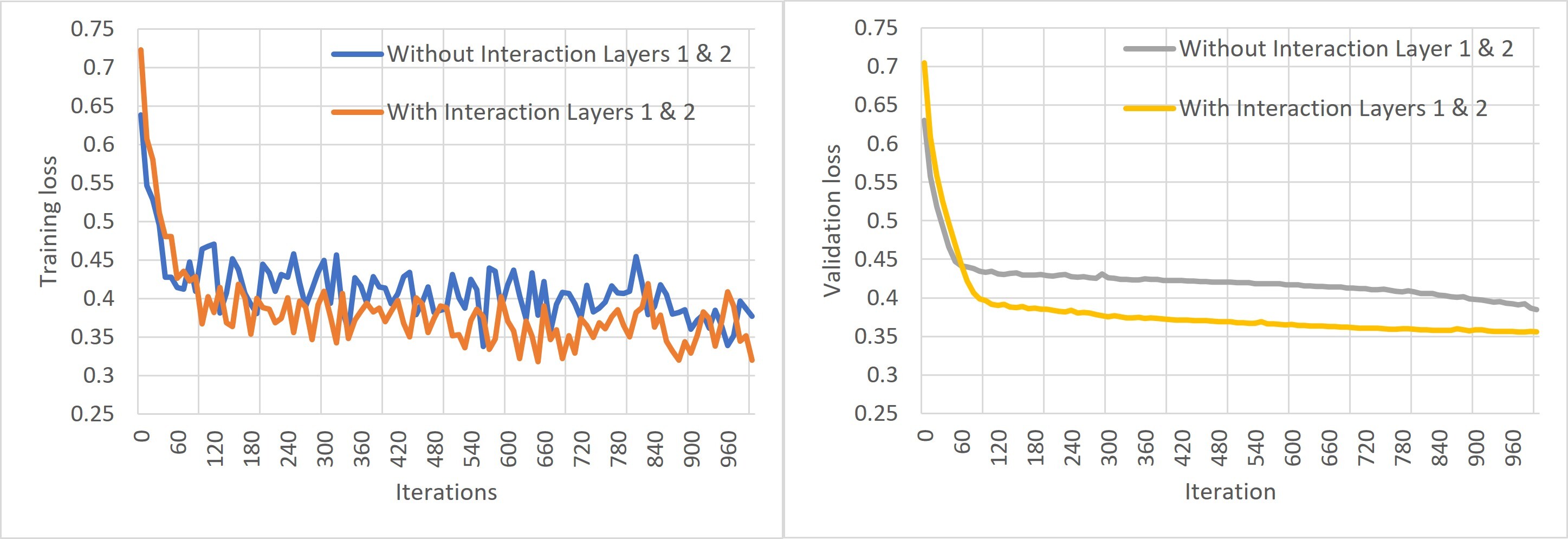}
\caption{Plots showing the loss on binary classification of training images (left) and testing images (right) of the \textit{Fashion MNIST} dataset over 1000 iterations using Ansatz 1, with and without Interaction Layers 1 and 2. Both show that when Interaction layers are present, the costs converge to a lower value. The convergence also occurs more rapidly when these layers are used.}
\label{fig:entang_cost}
\end{figure}
  \hfill
  \begin{figure}[!ht]
    \centering
    \includegraphics[width=8cm]{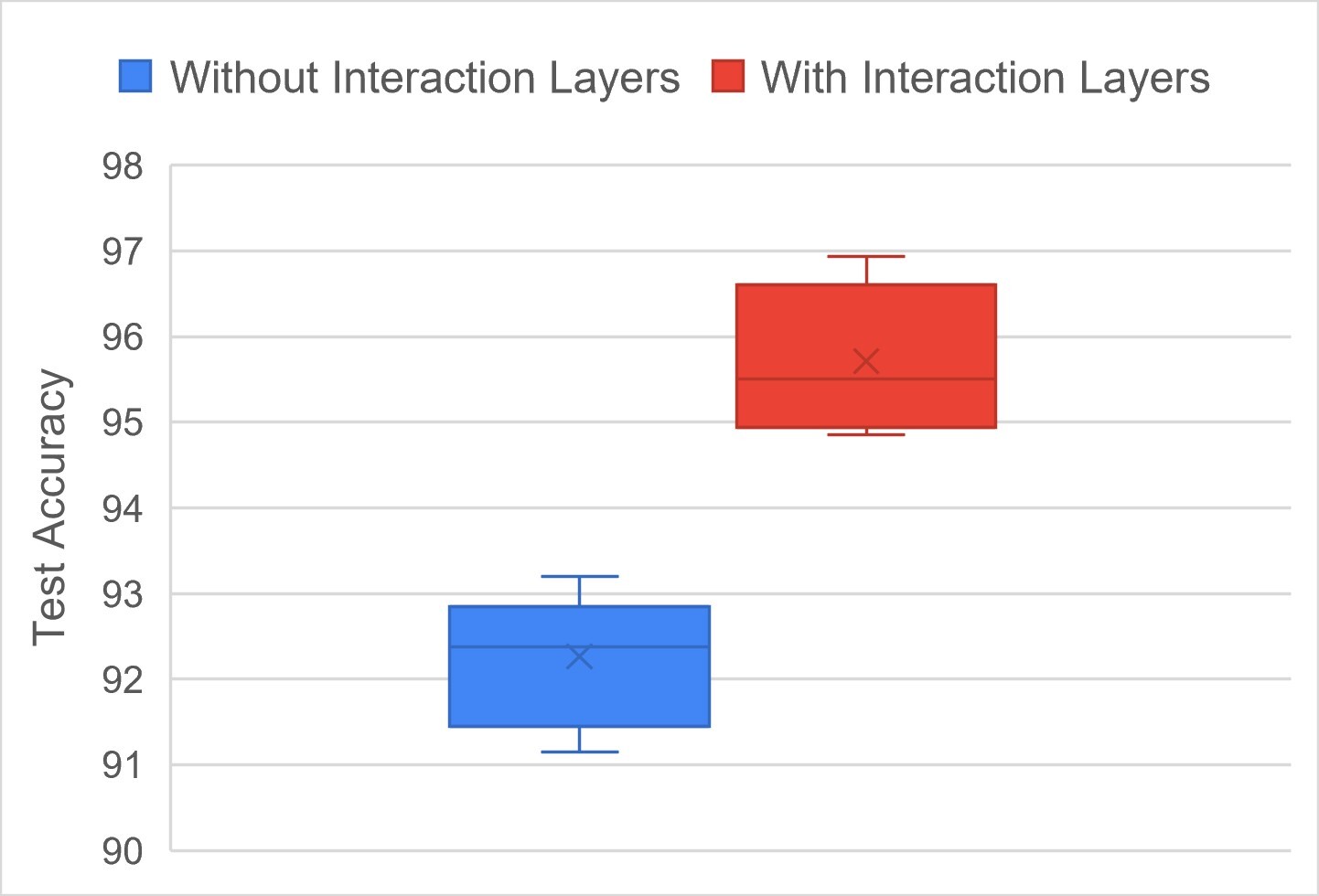}
    \caption{Results showing test accuracies in the presence and absence of Interaction Layers 1 and 2 in our proposed architecture. The results are shown for binary classification of the \textit{Fashion MNIST} dataset using Ansatz 1 and \textit{Angle Encoding}. It shows clear superiority in classifier performance when these novel layers are present. The results have been computed over 10 independent runs.}
    \label{fig:entang_acc}
  \end{figure}

The superiority in performance of ansatz 1 over ansatz 2, due to its additional parameters, is demonstrated in Fig. \ref{fig:acc_angvsamp}, where the range of testing (validation) accuracies are summarized. It can be noticed that for the same encoding scheme, ansatz 1 outperforms ansatz 2. It must be noted that Fig. \ref{fig:acc_angvsamp} is drawn for the \textit{Fashion MNIST} dataset, and hence, in both cases of ansatzes \textit{Angle Encoding} outperforms \textit{Amplitude Encoding}. 

It can be observed from table \ref{tab:1} that the type of Quantum Encoding method used to bear the best accuracies is dependent on the dataset that is used. The peak accuracy for the \textit{Fashion MNIST} dataset results from reducing the classical data by an autoencoder followed by \textit{Angle Encoding} whereas, for \textit{MNIST}, it is simple resizing followed by \textit{Amplitude Encoding}. Comparison of the results with other existing quantum machine learning models for binary classification, such as that proposed in \cite{tak_hur_boss}, shows that our model surpasses their accuracy, as shown in the first two rows of table \ref{tab:2}. The mean accuracy for the binary classification of classes 0 and 1 for \textit{Fashion MNIST} is 1.45\%, and for \textit{MNIST}, it is 0.3\% more than that reported in \cite{tak_hur_boss}. This increase in accuracy can be attributed to the incorporation of the Interaction Layers and the use of the ancilla-based Classifier. It is observed that the proposed network shows a more minor standard deviation compared to that reported at \cite{tak_hur_boss}, which means that it is less sensitive to random initializations. \cite{tak_hur_boss} have further shown that their quantum network outperforms classical counterparts using a similar number of trainable parameters for the binary classification problem. 
It can, therefore, be concluded that the results of the study in our paper exhibit a clear superiority in performance compared to the classical networks with a similar number of trainable parameters. 
The peak accuracy for ansatz 1 used to construct the Convolutional Layers is $95.75\%\pm0.80\%$ compared to $96.84\%\pm1.50\%$ when ansatz 2 is used. This increase can be related to the number of trainable parameters available for each ansatz, which is higher in the case of ansatz 1. 
\begin{table}[t]
    \caption{Table showing a comparison of the results of our proposed model to that of existing models for different datasets.} 
    \begin{tabular}{cccc}
    \hline
    \textbf{No. of classes} & \textbf{Dataset}               & \textbf{Model used}  & \textbf{Test Accuracy (\%)} \\ \hline
    \multirow{3}{*}{$2$} & \multirow{3}{*}{Fashion MNIST} & Proposed    & $95.75\pm{0.80}$\\  
                                   & & QCNNFCDC (\cite{tak_hur_boss})       & $94.30\pm{1.60}$  
                \\      & & Proposed without E & $92.27\pm{0.69}$ \\ 
                                   \hline
    \multirow{3}{*}{$2$} & \multirow{2}{*}{MNIST}         & Proposed            & $99.00\pm{0.91}$                  \\  
                                  & & QCNNFCDC (\cite{tak_hur_boss})            & $98.70\pm{2.4}$                 \\      & & Proposed without E & $98.76\pm{0.35}$  \\
                                   \hline
    \multirow{2}{*}{$3$} & \multirow{2}{*}{Iris}         & Proposed            & $94.214\pm{1.11}$                 \\  
                                 &  & HCQAMC (\cite{chalumuri})            & $92.10$                 \\ 
                                   \hline
     
    \end{tabular}
\footnotetext{Note: `E' refers to Interaction Layers 1 and 2.}
\label{tab:2}
\end{table}
Additionally, the effect of the proposed Interaction Layers 1 and 2 on the performance of the network is demonstrated by comparing the performances with and without their presence. As evident in Fig. \ref{fig:entang_cost} and Fig. \ref{fig:entang_acc}, it can be concluded that these layers help reduce cost and increase accuracy by creating further dependencies between quantum states and making them more capable of spanning the \textit{Hilbert Space}. In both cases, the data reduction and quantum encoding technique used is autoencoder and \textit{Angle Encoding} respectively on the \textit{Fashion MNIST} dataset with ansatz 1 in Fig. \ref{fig:ansatz} used as the convolutional filter.

\subsubsection{Multiclass classification} \label{sec:multiclass_results}
 \begin{figure}[b]
    \centering
    \includegraphics[scale=1.4]{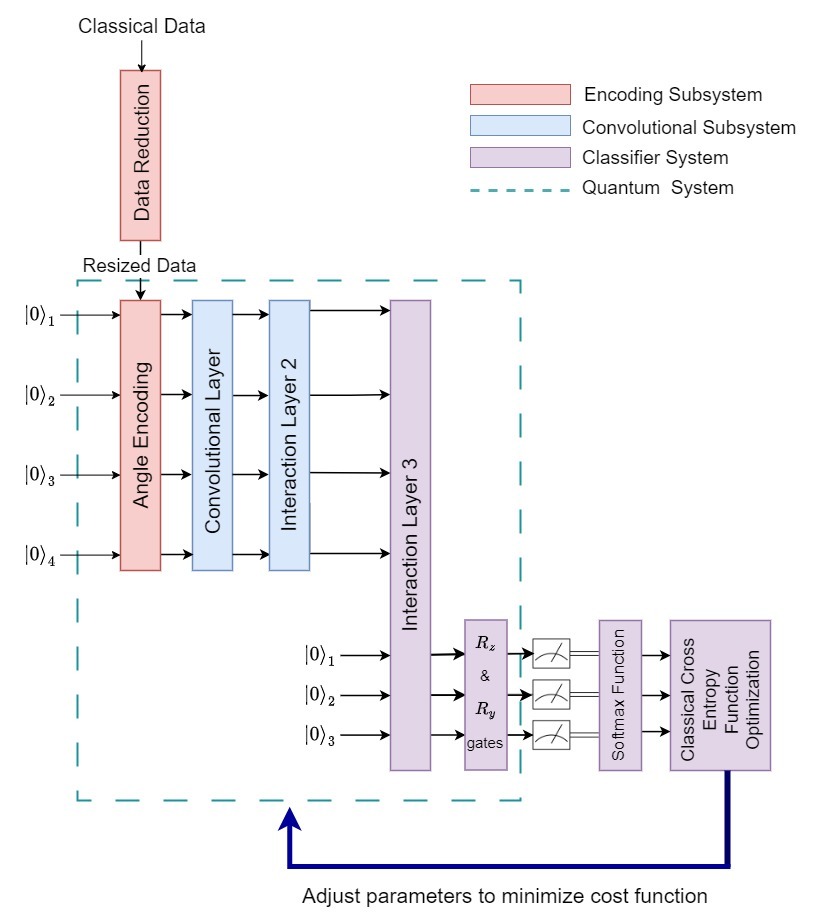}
    \caption{The proposed architecture for classifying the \textit{Iris dataset} comprises only four qubits as the four features can be embedded sufficiently using \textit{Angle Encoding}. Due to the small size of the network, a pooling layer is not required. Three ancilla qubits are used in the Classifier subsystem to classify the three classes.}
    \label{fig:iris_struc}
  \end{figure}
Multiclass classification is performed on the \textit{MNIST} and \textit{Fashion MNIST} datasets with the network slightly modified to include two Convolutional Layers cascaded together in each convolutional stage. It must be noted that these cascaded Convolutional Layers share the same weight. Therefore, the number of trainable parameters in the circuit does not significantly increase (the only increase can be attributed to the $3$ additional gates $R_x, R_y, R_z$ on the additional ancilla qubit related to the new class). The number and placement of the Interaction Layers remain unchanged from the network for Binary classification. Classes 0, 1, and 2 are selected for both datasets, and the batch size is kept at 100 with the learning rate set at 0.05 in the beginning and adapted to 0.01 after 50 iterations. 
The peak classification accuracy obtained is $91.53\%\pm0.98\%$ for the \textit{Fashion MNIST} dataset and $90.05\%\pm2.07\%$ for the \textit{MNIST} dataset using ansatz 1. The total number of trainable parameters in the network is only 53.

It is noticed that the combination of ansatz 1 and \textit{Angle Encoding} as the Quantum Encoding Method provides the highest accuracy for the \textit{Fashion MNIST} dataset, but the combination of ansatz 1 with the \textit{Amplitude Encoding} dataset performs better for \textit{MNIST} dataset. 

\begin{table}[]
\centering
\caption{Results of the Multiclass Classification problems different datasets. For each dataset, three classes have been classified. The results are consistent with that of binary classification- for \textit{Fashion MNIST}, \textit{Angle Encoding} performs better, and for \textit{MNIST}, \textit{Amplitude Encoding} gives higher accuracy. Autoencoder and Resize were used as data preprocessing methods for Angle and \textit{Amplitude Encoding} methods, respectively. }
\label{tab:my-table}
\begin{tabular}{cccc}
\hline
\textbf{Dataset}       & \textbf{Encoding Method} & \textbf{Trainable Parameters} & \textbf{Accuracy(\%)} \\ \hline
\multirow{2}{*}{Fashion MNIST}  & Angle           & \multirow{2}{*}{53}                   & $91.53\pm0.98$\%  \\
& Amplitude & & $89.85\pm1.20$\%\\\hline
\multirow{2}{*}{MNIST}  & Angle           & \multirow{2}{*}{53}                   & $88.52\pm0.93$\%\\
& Amplitude & & $90.05\pm2.07$\%\\\hline
\multirow{1}{*}{Iris}  & Angle           & \multirow{1}{*}{31}                   & $94.21\pm1.11$\%  \\\hline
\end{tabular}
\footnotetext{Note: In all of the cases, ansatz 1 is used to construct the Convolutional Layers.}
\label{tab:3}
\end{table}

To demonstrate the flexibility of the proposed circuit, performance on the \textit{Iris} dataset is also tested. In order to accommodate data of smaller dimensions, a cut-down version of the proposed circuit, with only 4 qubits, was sufficient. The reduced structure is visualized in Fig. \ref{fig:iris_struc}. The test accuracy with a batch size of 50 and a learning rate of 0.005 is found to be $94.21\%\pm1.11\%$. A three-class classifier with a variational quantum circuit has been proposed in \cite{chalumuri}, where classification was performed on classical one-dimensional feature data. The accuracy of $94.21\%\pm1.11\%$ supersedes the accuracy of the network proposed in \cite{chalumuri} (92.10\%) as shown in the third row of table \ref{tab:2}. It must also be noted that the network used for benchmarking the \textit{Iris} dataset has only 31 parameters and is much shallower than the one in \cite{chalumuri}. It is, therefore, understood that the network is not only limited to image classification but performs equally well in one-dimensional feature data.
The results of Multiclass classification problems are summarized in \ref{tab:3}.

The high accuracy achieved with only 53 parameters (for \textit{Fashion MNIST} and \textit{MNIST}) and 31 parameters (for \textit{Iris}) can be directly attributed to the incorporation of the Interaction Layers. When expanded qubit interactions are used, it is possible to achieve such accuracy while using a few parameters. This implies that these interactions will speed up the training of QCNNs while producing better outcomes with shallower circuits, enabling the development of networks that are more resistant to the barren plateau issues that result from greater depth. 

\section{Conclusion}
In this work, a shallow entangled QCNN with a minimal number of trainable parameters is proposed, which provides very satisfactory performance in binary and multiclass classification problems. The incorporation of weighted Interaction Layers, consisting of trainable parameters and utilizing three-qubit interactions between the quantum Convolutional Layers, and the use of the ancilla-based Classifier have played a significant role in enhancing the performance of the network. In doing so, it also studies the effect of the addition of such parameterized 3-qubit layers in a QCNN structure, which is a first of its kind. This result indicates the significance of increased qubit interaction on the substantial increase in the ability of a quantum network to learn more complex information from the training data while only using a few parameters. 

This approach constitutes a novel way towards the development of a generalized parameterized QNN that performs equally well for binary and multiclass classification on both image data and one-dimensional feature data. It further explores the possibilities of performance enhancement of quantum networks upon the use of increased qubit interaction, which is expected to be a reality in the not-so-distant future.

The simulation results indicate several advantages of the network, showing a clear superiority in performance compared to its counterparts using a similar number of parameters. Further research could be conducted to gain a more comprehensive understanding of the quantum advantage of these networks. An extensive investigation of the underlying causes of data dependencies on the feature encoding methods can be done. Other future milestones may also include an extension of the work for big data analysis and the solution of more complex problems utilizing more resources and power on real quantum computers.

\section*{Data Availability}
The simulation code used in this paper can be found at \href{https://github.com/chacconed/Quantum-Convolutional-Neural-Networks-with-Interaction-Layers-for-Classification-of-Classical-Data}{Simulation Code}. The datasets used in this paper are publicly available and can be found in the works of \cite{deng2012mnist}, \cite{xiao2017fashion}, and \cite{de2015mobile}.

\section*{Statements and Declarations}
The authors declare no competing interest in any other work or publication.

\subsection*{Authors' contributions}
J. Mahmud worked on technical writing, coding, analyzing, and designing. R. Mashtura worked on technical writing, designing, and producing data and figures. S. A. Fattah was involved in technical writing, analysis, and design. M. Saquib participated in technical writing and idea exchange. 

\subsection*{Funding}
No funding has been received for this research work.

\bibliography{sn-bibliography}

\begin{thebibliography}{32}
\providecommand{\natexlab}[1]{#1}
\providecommand{\url}[1]{{#1}}
\providecommand{\urlprefix}{URL }
\providecommand{\doi}[1]{\url{https://doi.org/#1}}
\providecommand{\eprint}[2][]{\url{#2}}
 \bibcommenthead

\bibitem[{Araujo et~al(2021)Araujo, Park, Petruccione, and da~Silva}]{div_conq}
Araujo IF, Park DK, Petruccione F, et~al (2021) A divide-and-conquer algorithm for quantum state preparation. Scientific Reports 11(1):1--12. \doi{https://doi.org/10.1038/s41598-021-85474-1}

\bibitem[{Arute et~al(2019)Arute, Arya, Babbush, Bacon, Bardin, Barends, Biswas, Boixo, Brandao, Buell et~al}]{r1}
Arute F, Arya K, Babbush R, et~al (2019) Quantum supremacy using a programmable superconducting processor. Nature 574(7779):505--510. \doi{https://doi.org/10.1038/s41586-019-1666-5}

\bibitem[{Ayoade et~al(2022)Ayoade, Rivas, and Orduz}]{ayoade2022artificial}
Ayoade O, Rivas P, Orduz J (2022) Artificial intelligence computing at the quantum level. Data 7(3):28

\bibitem[{Bergholm et~al(2018)Bergholm, Izaac, Schuld, Gogolin, Ahmed, Ajith, Alam, Alonso-Linaje, AkashNarayanan, Asadi et~al}]{bergholm2018pennylane}
Bergholm V, Izaac J, Schuld M, et~al (2018) Pennylane: Automatic differentiation of hybrid quantum-classical computations. arXiv preprint arXiv:181104968 \doi{https://doi.org/10.48550/arXiv.1811.04968}

\bibitem[{Boyd and Vandenberghe(2004)}]{boyd2004convex}
Boyd SP, Vandenberghe L (2004) Convex optimization. Cambridge university press

\bibitem[{Chalumuri et~al(2021)Chalumuri, Kune, and Manoj}]{chalumuri}
Chalumuri A, Kune R, Manoj B (2021) A hybrid classical-quantum approach for multi-class classification. Quantum Information Processing 20(3):1--19. \doi{https://doi.org/10.1007/s11128-021-03029-9}

\bibitem[{Cong et~al(2019)Cong, Choi, and Lukin}]{cong_quantum_org}
Cong I, Choi S, Lukin MD (2019) Quantum convolutional neural networks. Nature Physics 15(12):1273--1278. \doi{https://doi.org/10.1038/s41567-019-0648-8}

\bibitem[{De~Marsico et~al(2015)De~Marsico, Nappi, Riccio, and Wechsler}]{de2015mobile}
De~Marsico M, Nappi M, Riccio D, et~al (2015) Mobile iris challenge evaluation (miche)-i, biometric iris dataset and protocols. Pattern Recognition Letters 57:17--23

\bibitem[{Deng(2012)}]{deng2012mnist}
Deng L (2012) The mnist database of handwritten digit images for machine learning research [best of the web]. IEEE signal processing magazine 29(6):141--142

\bibitem[{Enos et~al(2021)Enos, Reagor, Henderson, Young, Horton, Birch, and Rigetti}]{weather}
Enos GR, Reagor MJ, Henderson MP, et~al (2021) Synthetic weather radar using hybrid quantum-classical machine learning. arXiv preprint arXiv:211115605 \doi{https://doi.org/10.48550/arXiv.2111.15605}

\bibitem[{Farhi and Neven(2018)}]{farhi2018classification}
Farhi E, Neven H (2018) Classification with quantum neural networks on near term processors. arXiv preprint arXiv:180206002

\bibitem[{Hur et~al(2022)Hur, Kim, and Park}]{tak_hur_boss}
Hur T, Kim L, Park DK (2022) Quantum convolutional neural network for classical data classification. Quantum Machine Intelligence 4(1):1--18. \doi{https://doi.org/10.1007/s42484-021-00061-x}

\bibitem[{Jain et~al(2020)Jain, Ziauddin, Leonchyk, Yenkanchi, and Geraci}]{lung_cancer}
Jain S, Ziauddin J, Leonchyk P, et~al (2020) Quantum and classical machine learning for the classification of non-small-cell lung cancer patients. Springer Nature Applied Sciences 2(6):1--10. \doi{https://doi.org/10.1007/s42452-020-2847-4}

\bibitem[{Kerenidis and Prakash(2022)}]{kerenidis2022quantum}
Kerenidis I, Prakash A (2022) Quantum machine learning with subspace states. arXiv preprint arXiv:220200054

\bibitem[{Liu et~al(2021)Liu, Lim, Wood, Huang, Guo, and Huang}]{liu_the2nd}
Liu J, Lim KH, Wood KL, et~al (2021) Hybrid quantum-classical convolutional neural networks. Science China Physics, Mechanics and Astronomy 64(9):1--8. \doi{https://doi.org/10.1007/s11433-021-1734-3}

\bibitem[{Madzik et~al(2022)Madzik, Asaad, Youssry, Joecker, Rudinger, Nielsen, Young, Proctor, Baczewski, Laucht et~al}]{mkadzik2022precision}
Madzik MT, Asaad S, Youssry A, et~al (2022) Precision tomography of a three-qubit donor quantum processor in silicon. Nature 601(7893):348--353. \doi{https://doi.org/10.1038/s41586-021-04292-7}

\bibitem[{Mengoni and Di~Pierro(2019)}]{mengoni_kernel}
Mengoni R, Di~Pierro A (2019) Kernel methods in quantum machine learning. Quantum Machine Intelligence 1(3):65--71. \doi{https://doi.org/10.1007/s42484-019-00007-4}

\bibitem[{Nesterov(1983)}]{nesterov}
Nesterov YE (1983) A method for solving the convex programming problem with convergence rate. In: Dokl. Akad. Nauk SSSR,, pp 543--547

\bibitem[{Nguyen(2023)}]{nguyen2023biomarker}
Nguyen N (2023) Biomarker discovery with quantum neural networks: A case-study in ctla4-activation pathways. arXiv preprint arXiv:230601745

\bibitem[{Nguyen and Chen(2022)}]{nguyen2022quantum}
Nguyen N, Chen KC (2022) Quantum embedding search for quantum machine learning. IEEE Access 10:41444--41456

\bibitem[{Pesah et~al(2021)Pesah, Cerezo, Wang, Volkoff, Sornborger, and Coles}]{pesah2021absence}
Pesah A, Cerezo M, Wang S, et~al (2021) Absence of barren plateaus in quantum convolutional neural networks. Physical Review X 11(4):041011. \doi{https://doi.org/10.1103/PhysRevX.11.041011}

\bibitem[{Rebentrost et~al(2014)Rebentrost, Mohseni, and Lloyd}]{QSVM_rebentrost}
Rebentrost P, Mohseni M, Lloyd S (2014) Quantum support vector machine for big data classification. Physical Review Letters 113(13):130503. \doi{10.1103/PhysRevLett.113.130503}

\bibitem[{Sabour et~al(2017)Sabour, Frosst, and Hinton}]{sabour2017dynamic}
Sabour S, Frosst N, Hinton GE (2017) Dynamic routing between capsules. Advances in neural information processing systems 30

\bibitem[{Schuld(2021)}]{schuld2021supervised}
Schuld M (2021) Supervised quantum machine learning models are kernel methods. arXiv preprint arXiv:210111020 \doi{https://doi.org/10.48550/arXiv.2101.11020}

\bibitem[{Schuld and Killoran(2022)}]{schuld2022quantum}
Schuld M, Killoran N (2022) Is quantum advantage the right goal for quantum machine learning? Prx Quantum 3(3):030101

\bibitem[{Schuld and Petruccione(2018)}]{schuld2018supervised}
Schuld M, Petruccione F (2018) Supervised learning with quantum computers, vol~17. Springer, \doi{https://doi.org/10.1007/978-3-319-96424-9}

\bibitem[{Schuld et~al(2020)Schuld, Bocharov, Svore, and Wiebe}]{ckt_centric}
Schuld M, Bocharov A, Svore KM, et~al (2020) Circuit-centric quantum classifiers. Phys Rev A 101:032308. \doi{10.1103/PhysRevA.101.032308}, \urlprefix\url{https://link.aps.org/doi/10.1103/PhysRevA.101.032308}

\bibitem[{Schuld et~al(2021)Schuld, Sweke, and Meyer}]{schuld2021effect}
Schuld M, Sweke R, Meyer JJ (2021) Effect of data encoding on the expressive power of variational quantum-machine-learning models. Physical Review A 103(3):032430

\bibitem[{Sim et~al(2019)Sim, Johnson, and Aspuru-Guzik}]{sim2019expressibility}
Sim S, Johnson PD, Aspuru-Guzik A (2019) Expressibility and entangling capability of parameterized quantum circuits for hybrid quantum-classical algorithms. Advanced Quantum Technologies 2(12):1900070

\bibitem[{Von~Lilienfeld(2018)}]{chemical}
Von~Lilienfeld OA (2018) Quantum machine learning in chemical compound space. Angewandte Chemie International Edition 57(16):4164--4169. \doi{https://doi.org/10.1002/anie.201709686}

\bibitem[{Wiebe et~al(2012)Wiebe, Braun, and Lloyd}]{Q_algo_fitting}
Wiebe N, Braun D, Lloyd S (2012) Quantum algorithm for data fitting. Phys Rev Lett 109:050505. \doi{10.1103/PhysRevLett.109.050505}, \urlprefix\url{https://link.aps.org/doi/10.1103/PhysRevLett.109.050505}

\bibitem[{Xiao et~al(2017)Xiao, Rasul, and Vollgraf}]{xiao2017fashion}
Xiao H, Rasul K, Vollgraf R (2017) Fashion-mnist: a novel image dataset for benchmarking machine learning algorithms. arXiv preprint arXiv:170807747

\end{thebibliography}

\clearpage


\section*{Appendix A: Relevant Quantum Gates}\label{secA1}
\begin{table*}[!ht]
    \caption{Summary of the quantum gates used in the architecture.}
    
    \centering
    \resizebox{\columnwidth}{!}{%
    \begin{tabular}{llll}
    \hline
        \textbf{Gates} & \textbf{Graph Representation} & \textbf{Graphical Form} & \textbf{Properties} \\ \hline\\
        Pauli-X &     \( 
    \Qcircuit @C=1em @R=.7em {
    & \gate{X} & \qw
    } 
    \) & $ \begin{bmatrix}
    0 & 1 \\
    1 & 0 
    \end{bmatrix}  $ & \makecell{Rotates a qubit by $180$\textdegree\\ about the x-axis} \\ \\\hline\\
        $R_x$ &     \( 
    \Qcircuit @C=1em @R=.7em {
    & \gate{R_x(\theta)} & \qw
    } 
    \) & $ \begin{bmatrix}
    \cos(\frac{\theta}{2}) & -i\sin(\frac{\theta}{2}) \\
    -i\sin(\frac{\theta}{2}) & \cos(\frac{\theta}{2}) 
    \end{bmatrix}  $ & \makecell{Rotates the qubit's state vector\\ about the x-axis by an angle $\theta$} \\ \\ \hline\\
        $R_y$ &     \( 
    \Qcircuit @C=1em @R=.7em {
    & \gate{R_y(\theta)} & \qw
    } 
    \) & $ \begin{bmatrix}
    \cos(\frac{\theta}{2}) & -\sin(\frac{\theta}{2}) \\
    \sin(\frac{\theta}{2}) & \cos(\frac{\theta}{2}) 
    \end{bmatrix}  $ & \makecell{Rotates the qubit's state vector\\ about the y-axis by an angle $\theta$} \\\\ \hline\\
        $R_z$ &     \( 
    \Qcircuit @C=1em @R=.7em {
        & \gate{R_z(\theta)} & \qw
    } \) & $ \begin{bmatrix} \text{e}^{-i\frac{\theta}{2}}  & 0 \\
    0 & \text{e}^{i\frac{\theta}{2}} \end{bmatrix} $ & \makecell{Rotates the qubit's state vector\\ about the z-axis by an angle $\theta$} \\ \\ \hline \\
        Controlled $R_x$ & \(\Qcircuit @C=1em @R=.7em {
     & \ctrl{1} & \qw \\
     & \gate{R_z} & \qw
}\) & $ \begin{bmatrix}
    1 & 0 & 0 & 0 \\
    0 & \cos(\frac{\theta}{2}) & 0 & -i\sin(\frac{\theta}{2}) \\
    0 & 0 & 1 & 0 \\
    0 & -i\sin(\frac{\theta}{2}) & 0 & \cos(\frac{\theta}{2}) 
    \end{bmatrix}  $ & \makecell{The $R_x$ gate acts on the target qubit\\ if the control qubit is in the state $\ket{1}$} \\ \\ \hline \\ 
        Controlled $R_y$ & \(\Qcircuit @C=1em @R=.7em {
     & \ctrl{1} & \qw \\
     & \gate{R_y} & \qw
}\) & $ \begin{bmatrix}
    1 & 0 & 0 & 0 \\
    0 & \cos(\frac{\theta}{2}) & 0 & -\sin(\frac{\theta}{2}) \\
    0 & 0 & 1 & 0 \\
    0 & \sin(\frac{\theta}{2}) & 0 & \cos(\frac{\theta}{2}) 
    \end{bmatrix}  $ & \makecell{The $R_y$ gate acts on the target qubit\\ if the control qubit is in the state $\ket{1}$} \\ \\\hline\\
        CNOT            & \(\Qcircuit @C=1em @R=.7em {
     & \ctrl{1} & \qw \\
     & \targ & \qw
}\) & $ \begin{bmatrix}
    1 & 0 & 0 & 0 \\
    0 & 1 & 0 & 0 \\
    0 & 0 & 0 & 1 \\
    0 & 0 & 1 & 0 
    \end{bmatrix}  $  & \makecell{The Pauli-X gate acts on the target qubit\\ if the control qubit is in the state \ket{1}} \\ \\ \hline \\
            Toffoli         & \(\Qcircuit @C=1em @R=.7em {
     & \ctrl{1} & \qw \\
     & \ctrl{1} & \qw \\
     & \targ & \qw
}\) & $ \begin{bmatrix}
    1 & 0 & 0 & 0 & 0 & 0 & 0 & 0 \\
    0 & 1 & 0 & 0 & 0 & 0 & 0 & 0 \\
    0 & 0 & 1 & 0 & 0 & 0 & 0 & 0 \\
    0 & 0 & 0 & 1 & 0 & 0 & 0 & 0 \\
    0 & 0 & 0 & 0 & 1 & 0 & 0 & 0 \\
    0 & 0 & 0 & 0 & 0 & 1 & 0 & 0 \\
    0 & 0 & 0 & 0 & 0 & 0 & 0 & 1 \\
    0 & 0 & 0 & 0 & 0 & 0 & 1 & 0 
    \end{bmatrix}  $  & \makecell{The target qubit is inverted if both\\the control qubits are \ket{1}} \\ \\ \hline \\   
        U3             & 
    \( 
    \Qcircuit @C=1em @R=.7em {
          & \gate{U3(\theta_1,\theta_2,\theta_3)} & \qw
    }
    \)  & $\begin{bmatrix} \cos(\frac{\theta_1}{2}) & -\text{e}^{i\theta_3}\sin(\frac{\theta_1}{2})  \\
    \text{e}^{i\theta_2}\sin(\frac{\theta_1}{2}) & \text{e}^{i(\theta_2+\theta_3)}\cos(\frac{\theta_1}{2}) \end{bmatrix}$  & \makecell{$\theta_1$ is the angle of rotation around \\the Bloch sphere's equator. $\theta_2$ and $\theta_3$\\ are the phase angle about the\\ z-axis and x-axis respectively} \\ \\ \hline
    \end{tabular}%
    }
    \label{tab:A}
\end{table*}
Note that the matrices shown in table \ref{tab:A} are for a composite state involving the number of qubits that the particular gate acts on. 

\clearpage
\section*{Appendix B: Model Weights}\label{secA1}
  \begin{figure}[!ht]
    \centering
    \includegraphics[width=\linewidth]{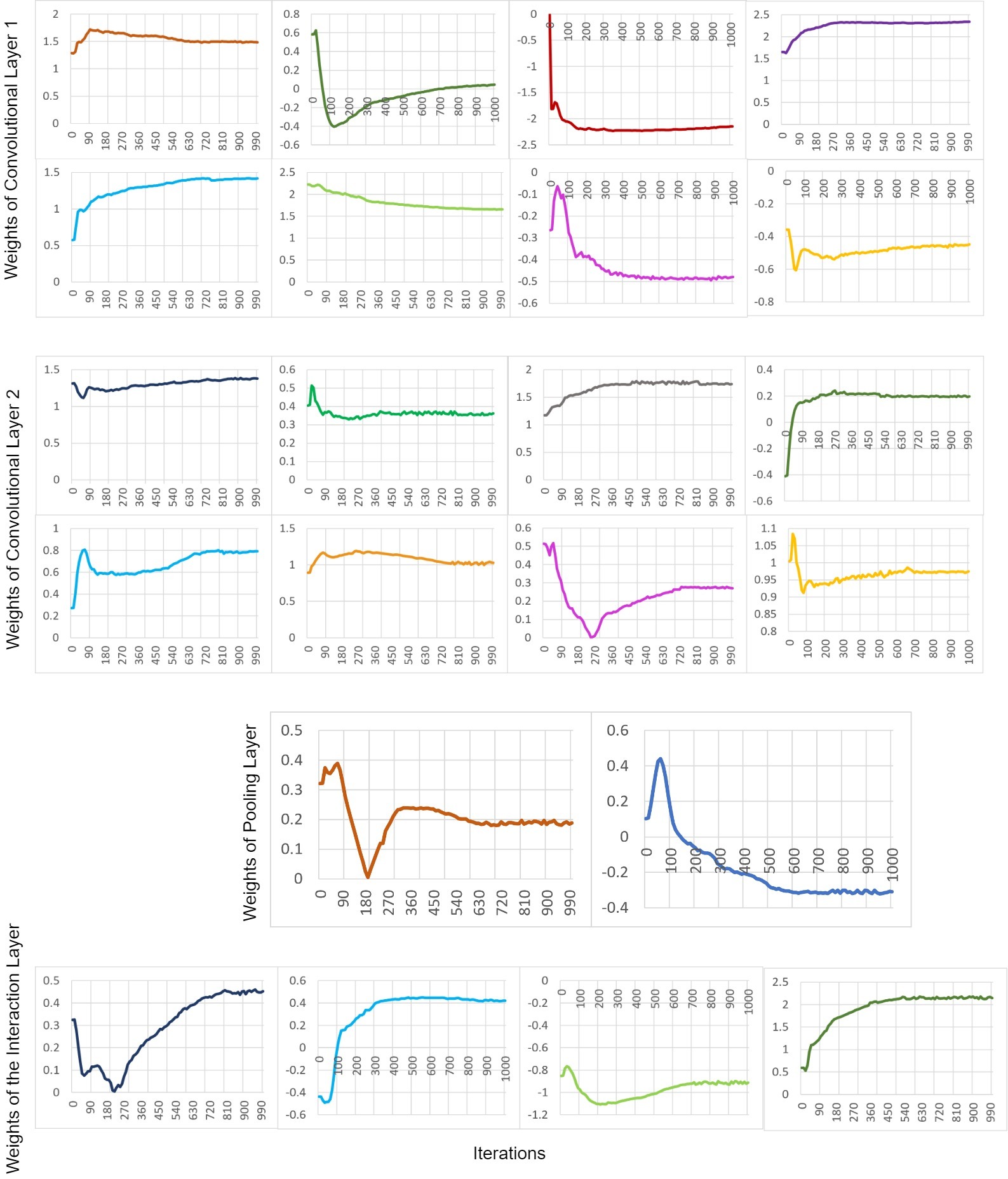}
    \caption{Learning curves of some of the parameters of the Convolutional Layer 1, 2, Pooling Layer, and Interaction Layers for binary classification of the \textit{Fashion MNIST} dataset using Ansatz 1 and \textit{Angle Encoding}. The weights start converging after 750 iterations, achieving an accuracy of $95.75\%\pm0.80$. }
    \label{fig:weights-conv1}
  \end{figure}


\end{document}